\documentclass[12pt]{article}
\pdfoutput=1
\usepackage{jheppub}
\usepackage{epsfig}
\usepackage{amsmath}
\usepackage{amssymb}
\usepackage{amsfonts}
\usepackage{amsxtra}
\usepackage{amsthm}
\usepackage{mathrsfs}
\usepackage{makeidx}
\usepackage{graphics}
\usepackage{dsfont}
\usepackage{mathtools}
\usepackage{graphicx}
\usepackage{subcaption}
\usepackage{placeins}
\usepackage{bm}
\usepackage[capitalise]{cleveref}
\usepackage{empheq}
\usepackage{colortbl}
\usepackage{xcolor}
\usepackage{enumerate}
\usepackage{titlesec}
\usepackage{longtable}
\usepackage{float}
\usepackage{color}
\usepackage{tikz}
\usepackage{xfrac}
\usepackage{footnote}
\usepackage{rotating}
\usepackage{lscape}
\usepackage{makecell}
\usepackage{environ}
\usepackage{tabularx}
\usepackage{subfiles}
\usepackage[export]{adjustbox}
\usepackage{ytableau}
\usepackage{tikz-3dplot}
\usepackage{slashed}
\usepackage{pifont}
\usepackage{multirow}
\usepackage{mdframed}
\usepackage{bbm}
\usepackage{ulem}

\usetikzlibrary{positioning,trees,decorations.pathmorphing,decorations.markings,decorations.pathreplacing,calc,shapes,patterns,arrows,chains,arrows.meta,fit,fadings,decorations.markings,graphs,graphs.standard,quotes}

\titleformat{\paragraph}
{\normalfont\normalsize\bfseries}{\theparagraph}{1em}{}
\titlespacing*{\paragraph}{0pt}{3.25ex plus 1ex minus .2ex}{1.5ex plus .2ex}

\DeclareMathOperator{\U}{U}
\DeclareMathOperator{\PSU}{PSU}
\DeclareMathOperator{\SU}{SU}
\DeclareMathOperator{\SO}{SO}

\DeclareMathOperator{\Pin}{Pin}
\DeclareMathOperator{\USp}{USp}

\def\CC{{\mathbb{C}}}

\theoremstyle{definition}

\catcode`\@=11   
\newdimen\@rotdimen
\newbox\@rotbox  

\def\@vspec#1{\special{ps:#1}}
\def\@rotstart#1{\@vspec{gsave currentpoint currentpoint translate
   #1 neg exch neg exch translate}}
\def\@rotfinish{\@vspec{currentpoint grestore moveto}}
\def\@rotr#1{\@rotdimen=\ht#1\advance\@rotdimen by\dp#1%
   \hbox to\@rotdimen{\hskip\ht#1\vbox to\wd#1{\@rotstart{90 rotate}%
   \box#1\vss}\hss}\@rotfinish}
\def\@rotl#1{\@rotdimen=\ht#1\advance\@rotdimen by\dp#1%
   \hbox to\@rotdimen{\vbox to\wd#1{\vskip\wd#1\@rotstart{270 rotate}%
   \box#1\vss}\hss}\@rotfinish}%
\def\@rotu#1{\@rotdimen=\ht#1\advance\@rotdimen by\dp#1%
   \hbox to\wd#1{\hskip\wd#1\vbox to\@rotdimen{\vskip\@rotdimen
   \@rotstart{-1 dup scale}\box#1\vss}\hss}\@rotfinish}%
\def\@rotf#1{\hbox to\wd#1{\hskip\wd#1\@rotstart{-1 1 scale}%
   \box#1\hss}\@rotfinish}%
\def\rotate{\@ifnextchar[{\@rotate}{\@rotate[l\right]}}
\def\@rotate[#1]#2{\setbox\@rotbox=\hbox{#2}\@nameuse{@rot#1}\@rotbox}

\catcode`\@=12

\usetikzlibrary{positioning}
\usetikzlibrary{chains}
\usetikzlibrary{arrows, arrows.meta ,fit,decorations.pathreplacing}
\tikzstyle{every picture}+=[remember picture]
\tikzstyle{na} = [baseline]

\usetikzlibrary{arrows, decorations.markings, calc, fadings, decorations.pathreplacing, patterns, decorations.pathmorphing, positioning}
\tikzset{>={Latex[width=1.5mm,length=1.5mm]}}
\usetikzlibrary{graphs,graphs.standard,quotes}

\def\node#1#2{\overset{#1}{\underset{#2}{{\color{gray} \bullet}}}}

\def\node#1#2{\overset{#1}{\underset{#2}{\circ}}}

\tikzstyle{every picture}+=[remember picture]
\tikzstyle{na} = [baseline=-.5ex]

\newcommand{\eg}{e.g.}

\newcommand{\ie}{i.e.}

\numberwithin{equation}{section}
\newcommand{\bes}[1]{\begin{equation} \begin{split} #1\end{split} \end{equation}}

\newcommand{\nn}{\nonumber}

\newcommand{\be}{\begin{equation}} \newcommand{\ee}{\end{equation}}
\newcommand{\bea}{\begin{equation} \begin{aligned}} \newcommand{\eea}{\end{aligned} \end{equation}}

\def\tilde{\widetilde}

\def\bar{\overline}

\def\rt2{\sqrt{2}}

\def\mod{{\rm mod}}

\def\CC{{\cal C}}

\def\CF{{\cal F}}

\def\CI{{\cal I}}

\def\CM{{\cal M}}
\def\CN{{\cal N}}

\def\CT{{\cal T}}

\def\CZ{{\cal Z}}

\def\1{{\ds 1}}

\def\SO{\mathrm{SO}}
\def\O{\mathrm{O}}
\def\SU{\mathrm{SU}}

\def\Spin{\mathrm{Spin}}

\def\su{\mathfrak{su}}

\def\so{\mathfrak{so}}

\def\usp{\mathfrak{usp}}

\def\repa{\raise4pt\hbox{$\square$}\mkern-14mu\raise-4pt\hbox{$\square$}}
\def\repab{\overline{\raise4pt\hbox{$\square$}\mkern-14mu\raise-4pt\hbox{$\square$}\mkern-1mu}}

\def\smileface{\ensuremath{\hbox{\large$\bigcirc$}\mkern-15mu\raise-1pt\hbox{\scriptsize$\smallsmile$}%
\mkern-10mu\raise4pt\hbox{..}\mkern4mu}}
\def\frownface{\ensuremath{\hbox{\large$\bigcirc$}\mkern-15mu\raise-1pt\hbox{\scriptsize$\smallfrown$}%
\mkern-10mu\raise4pt\hbox{..}\mkern4mu}}

\newcommand{\ba}{\begin{array}}
\newcommand{\ea}{\end{array}}
\newcommand{\bi}{\begin{itemize}}
\newcommand{\ei}{\end{itemize}}
\def\vec#1{\bm{#1}}
\def\bea#1\eea{\allowdisplaybreaks \begin{align}#1\end{align}}
 \newcommand{\ben}{\begin{enumerate}}
\newcommand{\een}{\end{enumerate}}
\newcommand{\bean}{\begin{eqnarray*}}
\newcommand{\eean}{\end{eqnarray*}}
\newcommand{\eref}[1]{(\ref{#1})}

\newcommand{\tQ}{\widetilde{Q}}

\newcommand{\BZ}{\mathbb{Z}}

\newcommand{\Sym}{\mathrm{Sym}}

\definecolor{light-gray}{gray}{0.5}

\newcommand{\blue}{\color{blue}}

\newcommand{\udl}[1]{\mathrm{d} #1 \,}

\def\ga{\alpha}

\def\aup#1 {\overset{#1}{\uparrow} \, \overset{\tilde{#1}}{\downarrow}}

\tikzset{snake it/.style={decorate, decoration={snake, amplitude=.4mm, segment length=2mm,
                       post length=0mm,pre length=0mm}}}
                       
 \newcommand{\GCD}{\mathrm{GCD}}

\hypersetup{
	pdftitle={Mixed Anomalies, Two-groups, Non-Invertible Symmetries, and 3d Superconformal Indices},    
	pdfauthor={\textcopyright\ Noppadol Mekareeya, Matteo Sacchi},     
	pdfsubject={hep-th},   
	pdfcreator={pdfLaTex},   
	pdfproducer={LaTex}, 
	pdfkeywords={},
	colorlinks=true,
}

\makeatletter
\newsavebox{\measure@tikzpicture}
\NewEnviron{scaletikzpicturetowidth}[1]{%
  \def\tikz@width{#1}%
  \begin{lrbox}{\measure@tikzpicture}%
  \BODY
  \end{lrbox}%
  \pgfmathparse{#1/\wd\measure@tikzpicture}%
  \BODY
}
\makeatother

\makeatletter
\def\squarecorner#1{
    \pgf@x=\the\wd\pgfnodeparttextbox%
    \pgfmathsetlength\pgf@xc{\pgfkeysvalueof{/pgf/inner xsep}}%
    \advance\pgf@x by 2\pgf@xc%
    \pgfmathsetlength\pgf@xb{\pgfkeysvalueof{/pgf/minimum width}}%
    \ifdim\pgf@x<\pgf@xb%
        \pgf@x=\pgf@xb%
    \fi%
    \pgf@y=\ht\pgfnodeparttextbox%
    \advance\pgf@y by\dp\pgfnodeparttextbox%
    \pgfmathsetlength\pgf@yc{\pgfkeysvalueof{/pgf/inner ysep}}%
    \advance\pgf@y by 2\pgf@yc%
    \pgfmathsetlength\pgf@yb{\pgfkeysvalueof{/pgf/minimum height}}%
    \ifdim\pgf@y<\pgf@yb%
        \pgf@y=\pgf@yb%
    \fi%
    \ifdim\pgf@x<\pgf@y%
        \pgf@x=\pgf@y%
    \else
        \pgf@y=\pgf@x%
    \fi
    \pgf@x=#1.5\pgf@x%
    \advance\pgf@x by.5\wd\pgfnodeparttextbox%
    \pgfmathsetlength\pgf@xa{\pgfkeysvalueof{/pgf/outer xsep}}%
    \advance\pgf@x by#1\pgf@xa%
    \pgf@y=#1.5\pgf@y%
    \advance\pgf@y by-.5\dp\pgfnodeparttextbox%
    \advance\pgf@y by.5\ht\pgfnodeparttextbox%
    \pgfmathsetlength\pgf@ya{\pgfkeysvalueof{/pgf/outer ysep}}%
    \advance\pgf@y by#1\pgf@ya%
}
\makeatother

\pgfdeclareshape{square}{
    \savedanchor\northeast{\squarecorner{}}
    \savedanchor\southwest{\squarecorner{-}}

    \foreach \x in {east,west} \foreach \y in {north,mid,base,south} {
        \inheritanchor[from=rectangle]{\y\space\x}
    }
    \foreach \x in {east,west,north,mid,base,south,center,text} {
        \inheritanchor[from=rectangle]{\x}
    }
    \inheritanchorborder[from=rectangle]
    \inheritbackgroundpath[from=rectangle]
}

\tikzset{stretch/.initial=1}
\newcommand\drawloop[4][]%
   {\draw[shorten <=0pt, shorten >=0pt,#1]
      ($(#2)!\pgfkeysvalueof{/tikz/stretch}!(#2.#3)$)
      let \p1=($(#2.center)!\pgfkeysvalueof{/tikz/stretch}!(#2.north)-(#2)$),
          \n1= {veclen(\x1,\y1)*sin(0.5*(#4-#3))/sin(0.5*(180-#4+#3))}
      in arc [start angle={#3-90}, end angle={#4+90}, radius=\n1]%
   }

\title{Mixed Anomalies, Two-groups, Non-Invertible Symmetries, and 3d Superconformal Indices}
\author[a,b]{Noppadol Mekareeya}
\author[c]{and Matteo Sacchi}
\affiliation[a]{INFN, sezione di Milano-Bicocca, \\Piazza della Scienza 3, I-20126 Milano, Italy}
\affiliation[b]{Department of Physics, Faculty of Science, \\
Chulalongkorn University, Phayathai Road, \\
Pathumwan, Bangkok 10330, Thailand}
\affiliation[c]{Mathematical Institute, University of Oxford, \\ Andrew-Wiles Building, Woodstock Road, \\
Oxford, OX2 6GG, United Kingdom}
\emailAdd{n.mekareeya@gmail.com}
\emailAdd{matteo.sacchi@maths.ox.ac.uk}
\abstract{Mixed anomalies, higher form symmetries, two-group symmetries and non-invertible symmetries have proved to be useful in providing non-trivial constraints on the dynamics of quantum field theories. We study mixed anomalies involving discrete zero-form global symmetries, and possibly a one-form symmetry, in 3d $\mathcal{N} \geq 3$ gauge theories using the superconformal index.  The effectiveness of this method is demonstrated via several classes of theories, including Chern-Simons-matter theories, such as the $\mathrm{U}(1)_k$ gauge theory with hypermultiplets of diverse charges, the $T(\SU(N))$ theory of Gaiotto-Witten, the theories with $\mathfrak{so}(2N)_{2k}$ gauge algebra and hypermultiplets in the vector representation, and variants of the Aharony-Bergman-Jafferis (ABJ) theory with the orthosymplectic gauge algebra. Gauging appropriate global symmetries of some of these models, we obtain various interesting theories with non-invertible symmetries or two-group structures.}
\begin{document}
\maketitle

\section{Introduction}
Symmetry serves as a vital organising principle in the study of quantum field theories. It can provide highly non-trivial constraints in the theory, for example, via selection rules and 't Hooft anomalies. There have been vast recent developments in this line of research.  One of the main important ideas is that a number of properties of the symmetries can be formulated in terms of the associated topological defects.  Specifically, if the symmetry obeys the group law, it can be viewed as the fusion rule of the topological defects as follows: the topological defects associated with the group elements $g$ and $h$ can be merged to form a topological defect associated with the group element $gh$.  

This point of view has led to a number of new concepts of generalised global symmetries. This includes {\it higher-form symmetries} \cite{Kapustin:2014gua, Gaiotto:2014kfa}\footnote{We remark that these have been worked out in many theories by means of defect groups, see \eg~ \cite{DelZotto:2015isa, Albertini:2020mdx, DelZotto:2020esg, Closset:2020scj, Closset:2021lwy, DelZotto:2022ras}.} whose topological defects have codimension greater than one and whose charged objects are extended operators, and {\it non-invertible symmetries} whose topological defects do not have an inverse and so do not form a group.  Examples of the latter in 2d theories and in 3d TQFTs have been known for some time, see for example \cite{Verlinde:1988sn,Petkova:2000ip,Fuchs:2002cm,Bachas:2004sy,Bachas:2009mc,Bhardwaj:2017xup,Lin:2022dhv,Komargodski:2020mxz,Tachikawa:2017gyf,Frohlich:2004ef,Frohlich:2006ch,Frohlich:2009gb,Carqueville:2012dk,Brunner:2013xna,Chang:2018iay,Lin:2019hks,Huang:2021zvu,Thorngren:2019iar,Thorngren:2021yso,Huang:2021nvb,Inamura:2022lun,Gaiotto:2020iye,Burbano:2021loy,Lu:2022ver,Kapustin:2010if,Kaidi:2021gbs}, but only very recently they have been studied in different field theories and especially in higher dimensions from many point of views, see \eg~ \cite{Heidenreich:2021xpr, Choi:2021kmx, Kaidi:2021xfk,Koide:2021zxj, Apruzzi:2021nmk, Cordova:2022rer,Benini:2022hzx,Hayashi:2022fkw, Kaidi:2022uux, Choi:2022zal, Bhardwaj:2022yxj,Roumpedakis:2022aik, Arias-Tamargo:2022nlf, Choi:2022jqy, Cordova:2022ieu,Antinucci:2022eat, Damia:2022rxw,Damia:2022bcd, Choi:2022rfe, Bashmakov:2022jtl, Bhardwaj:2022lsg,Lin:2022xod,Bartsch:2022mpm, Apruzzi:2022rei, GarciaEtxebarria:2022vzq, Heckman:2022muc, Freed:2022qnc, Kaidi:2022cpf,Niro:2022ctq,Antinucci:2022vyk} and \cite{Argyres:2022mnu,Cordova:2022ruw} for recent reviews. Yet another important idea which is central to this paper is the coexistence of a zero-form and a one-form global symmetry.  This can happen in several ways, for example, they can form a direct or a semi-direct product, there can be a mixed anomaly between them, or they can combine to form a non-trivial extension, where the latter is known as a {\it two-group symmetry} \cite{Sharpe:2015mja, Tachikawa:2017gyf, Benini:2018reh, Hsin:2020nts, Cordova:2020tij}. In this paper, we focus on the two-group symmetries that involve a discrete one-form symmetry and a continuous zero-form symmetry. This type of symmetry has been studied in a wide range of theories, see \eg~ \cite{DelZotto:2020sop, Hsin:2020nts,Apruzzi:2021vcu, Bhardwaj:2021wif, Lee:2021crt, Apruzzi:2021mlh, Genolini:2022mpi, DelZotto:2022fnw, DelZotto:2022joo,Cvetic:2022imb, Bhardwaj:2022scy, Carta:2022fxc,DelZotto:2022ohj}.

In this paper, we study mixed anomalies in three-dimensional superconformal field theories with $\CN \geq 3$ supersymmetry using the superconformal index \cite{Bhattacharya:2008zy,Bhattacharya:2008bja, Kim:2009wb,Imamura:2011su, Kapustin:2011jm, Dimofte:2011py} as a main tool.  This provides a convenient and efficient way to detect various mixed anomalies, including those involving two discrete zero-form global symmetries and a continuous zero-form flavour symmetries.  Specifically, motivated by \cite{Bhardwaj:2022dyt}, we calculate the index in a particular way in order to study the monopole operators carrying fractional magnetic fluxes for both the gauge group and the Cartan subalgebra of the flavour symmetry group, whose existence might signal the presence of the anomaly. From the perspective of the index, this is manifest in the fact that certain gaugings of the global symmetries are not allowed.

We demonstrate these ideas in the context of several Chern-Simons-matter theories, which include the $\U(1)_k$ gauge theory\footnote{In this paper, $G_k$ denotes gauge group or gauge algebra $G$ with Chern-Simons coefficient $k$.} with hypermultiplets with arbitrary charge, the theories with $\so(2N)_k$ gauge algebra and hypermultiplets in the vector representation, and several variants of the Aharony-Bergman-Jafferis (ABJ) theories \cite{Aharony:2008gk} with the orthosymplectic gauge algebra.  Among a number of these theories, we find that gauging a discrete zero-form global symmetry leads to a dual one-form symmetry that forms a two-group structure with the zero-form flavour symmetry \cite{Tachikawa:2017gyf}. We also study discrete mixed anomalies for the $T(\SU(N))$ theory of Gaiotto and Witten \cite{Gaiotto:2008ak} and show as an application of these how they can be used to recover some known facts about the global form of the global symmetries of the 4d $\mathcal{N}=2$ theories of class $\mathcal{S}$ \cite{Gaiotto:2009we} from the 3d mirror perspective.

Since it is going to play a crucial role in our discussion, let us briefly review the argument of \cite{Benini:2018reh} for why gauging a discrete symmetry with a suitable mixed anomaly gives a theory with a two-group symmetry. Suppose that we have a $d$-dimensional theory $\mathcal{T}$ on a manifold $X_d$ with an anomaly encoded in the anomaly theory defined on $Y_{d+1}$ such that $\partial Y_{d+1}=X_d$
\bes{\label{genanomaly}
\exp\left(\frac{2\pi i}{N}\int_{Y_{d+1}}A_{p+1}\cup \Theta\right)~,
}
where $A_{p+1}\in H^{p+1}(X_d,\mathbb{Z}_N^{[p]})$ is a background field for a $\mathbb{Z}_N^{[p]}$ $p$-form symmetry and $\Theta$ is some class valued modulo $N$ constructed from the background fields for some other global symmetries. The full partition function of the theory including the anomaly theory is schematically
\bes{
Z_{\mathcal{T}}[A_{p+1}]=\exp\left(\frac{2\pi i}{N}\int_{Y_{d+1}}A_{p+1}\cup \Theta\right)\int\mathcal{D}\Phi\exp\left(iS[\Phi,A_{p+1}]\right)~.
}
When we gauge the symmetry $\mathbb{Z}_N^{[p]}$ we promote $A_{p+1}$ to a dynamical field $a_{p+1}$ over which we sum in the partition function and we introduce a $(d-p-1)$-cochain $B_{d-p-1}$ which is the background field for the dual $\mathbb{Z}_N^{[d-p-2]}$ $(d-p-2)$-form symmetry and which couples to $a_{p+1}$ in the partition function
\bes{
Z_{\mathcal{T}/\mathbb{Z}_N^{[p]}}[B_{d-p-1}]&=\sum_{a_{p+1}\in H^{p+1}(X_d,\mathbb{Z}_N^{[p]})}\exp\left(\frac{2\pi i}{N}\int_{X_d}a_{p+1}\cup B_{d-p-1}\right)Z_{\mathcal{T}}[a_{p+1}]\\
&=\sum_{a_{p+1}\in H^{p+1}(X_d,\mathbb{Z}_N^{[p]})}\exp\left(\frac{2\pi i}{N}\left(\int_{X_d}a_{p+1}\cup B_{d-p-1}+\int_{Y_{d+1}}a_{p+1}\cup \Theta\right)\right)\\
&\qquad\times\int\mathcal{D}\Phi\exp\left(iS[\Phi,a_{p+1}]\right)~.
}
We can now extend the coupling $\int_{X_d}a_{p+1}\cup B_{d-p-1}$ to the bulk $Y_{d+1}$ and exploit the fact that $a_{p+1}$ is a $(p+1)$-cocycle $\delta a_{p+1}=0$ to rewrite the partition function as
\bes{
Z_{\mathcal{T}/\mathbb{Z}_N^{[p]}}[B_{d-p-1}]&=\sum_{a_{p+1}\in H^{p+1}(X_d,\mathbb{Z}_N^{[p]})}\exp\left(\frac{2\pi i}{N}\int_{Y_{d+1}}a_{p+1}\cup \left(\delta B_{d-p-1}+ \Theta\right)\right)\\
&\qquad\times\int\mathcal{D}\Phi\exp\left(iS[\Phi,a_{p+1}]\right)~.
}
The exponential factor is a gauge anomaly for the gauged $\mathbb{Z}^{[p]}$ symmetry, so requiring the partition function of the theory $\mathcal{T}/\mathbb{Z}_N^{[p]}$ to be well-defined we get
$\delta B_{d-p-1}+ \Theta=0\text{ mod }N$, or in other words in presence of non-trivial background fields for the other symmetries which appear in $\Theta$ we have that $B_{d-p-1}$ is not closed and instead
\bes{
\delta B_{d-p-1}= \Theta~.
}
This means that after gauging a $\mathbb{Z}_N^{[p]}$ $p$-form symmetry with the anomaly \eqref{genanomaly} we obtain a dual $\mathbb{Z}_N^{[d-p-2]}$ $(d-p-2)$-form symmetry which forms a two-group with Postnikov class $\Theta$ with the other symmetries.

The superconformal index in three dimensions is also sensitive to one-form symmetries, a fact that was already exploited for example in \cite{Eckhard:2019jgg,Beratto:2021xmn}. In particular, it depends on the global structure of the gauge group thorugh the summation over monopole sectors, which can be changed when gauging a one-form symmetry. Thanks to this, the index can also be used in an indirect way to detect anomalies involving a one-form symmetry, and we will be particularly interested in the mixed anomalies between two discrete zero-form symmetries and a one-form symmetry.  By indirect, we mean that we study the theory resulting from gauging the one-form symmetry which in three spacetime dimensions gives rise to a zero-form symmetry. In some cases we can see from the index of the resulting theory that the additional gauging of a zero-form symmetry is obstructed, thus indicating the presence of the mixed anomaly in the original theory.  As it was pointed out in \cite{Kaidi:2021xfk, Kaidi:2022uux, Bhardwaj:2022yxj}, if the resulting anomaly takes a suitable form, then it leads to interesting consequences after gauging.  In particular, if we gauge the two zero-form symmetries in the original theory, the remaining one-form symmetry is non-invertible.  On the other hand, if we gauge the one-form symmetry and one of the two zero-form symmetries, the other zero-form symmetry is non-invertible.  We investigate this in the context of the 3d $\CN = 3$ theories with $\so(2N)_{2k}$ gauge algebra with adjoint hypermultiplets, and the ABJ theories with the orthosymplectic gauge algebra.  In particular, for $k$ even, we find that the $\Pin(2N)_{2k}\times \USp(2M)_{-k}$ variant\footnote{In this paper, we use the same nomenclature of \cite{Aharony:2013kma} and use $\Pin$ to denote the theory obtained by straightforwardly gauging both the magnetic and the charge conjugation symmetry of the $\SO$ theory. This symmetry is also called $\Pin^+$ as opposed to another possible variant $\Pin^-$, especially in non-supersymmetric set-ups, see for example \cite{Cordova:2017vab}.} of the ABJ theory has a non-invertible one-form symmetry, whereas the $(\Spin(2N)_{2k}\times \USp(2M)_{-k})/\mathbb{Z}_2$ and the $(\O(2N)_{2k}\times \USp(2M)_{-k})/\mathbb{Z}_2$ variants of the ABJ theory have a non-invertible zero-form symmetry.

The paper is organised as follows. In Section \ref{sec:u1}, we study the $\CN=4$ $\U(1)$ gauge theory with $2$ hypermultiplets of charge $q$. This part serves as an introduction on how to use the superconformal index to detect mixed anomalies and it contains some results that were already known in the literature.  We also present a mirror theory for the $\U(1)$ gauge theory with $2$ hypermultiplets of charge $2$, which is isomorphic to the $\Spin(2)$ gauge theory with $1$ hypermultiplet in the vector representation, in this section.  In Section \ref{sec:tsun}, we generalise these results to the $T[\SU(N)]$ theory, showing that it also possesses a mixed anomaly and how this can be used to understand the global symmetry group of the models obtained by gauging various copies of it, such as the 3d mirrors of 4d class $\mathcal{S}$ theories.  In Section \ref{sec:so}, we consider the 3d $\CN=3$ theories with $\so(2N)_k$ gauge algebra and $N_f$ hypermultiplets in the vector representation. We discuss the mixed anomalies as well as the two-group structures in these theories in detail.  We also extend the results of Section \ref{sec:u1} to the 3d $\CN=3$ $\U(1)_k$ gauge theory with $N_f$ hypermultiplets of charge $q$ in Subsection \ref{sec:3dN3U1k}.  In Section \ref{sec:noninvertibleABJ}, we investigate mixed anomalies involving a one-form symmetry and two zero-form symmetries as well as the non-invertible symmetries in the theories with $\so(2N)_{2k}$ gauge algebra and $N_f$ adjoint hypermultiplets and the ABJ theory and their variants.  Finally, in Section \ref{sec:twogroupsABJ}, we explore the mixed anomalies between two discrete zero-form symmetries and the continuous zero-form flavour symmetry as well as the two-group symmetries in the variants ABJ theory with the orthosymplectic gauge algebra. We conclude in Section \ref{sec:concl} mentioning some open questions and possible future developments. In Appendix \ref{app:index} we review the aspects of the 3d supersymmetric index that are relevant for our discussion.

\section{$\CN=4$ $\U(1)$ gauge theory with $2$ hypermultiplets of charge $q$}
\label{sec:u1}
Let us consider the 3d $\CN=4$ $\U(1)$ gauge theory with $2$ hypermultiplets of charge $q$. For convenience, we denote this theory by $\CT_q$.  The index of this theory is given by (see Appendix \ref{app:index} for more details on our conventions)
\bes{ \label{indTq}
\CI_{\CT_q}(w, n| f, m; x) &= \sum_{l \in \BZ +\epsilon(m)} \CF_{\CT_q} (w, n| f, m| l; x)~, \\
\CF_{\CT_q} (w, n| f, m| l; x) &\equiv w^{l} \oint \frac{dz}{2\pi i z} z^{n}~ \prod_{s=\pm1} \CI^{\frac{1}{2}}_\chi \left((z^q f)^{s}; s (q l + m); x \right) \times  \\
& \qquad \qquad     \CI^{\frac{1}{2}}_\chi \left((z^{-q} f)^{s}; s (-q l + m); x \right)~, 
}
where $w, n$ are the fugacity and background flux associated with the topological (magnetic) zero-form symmetry and $f, m$ are the fugacity and background flux associated with the flavour zero-form symmetry.  Note that $m$ and $n$ can be integral or half-odd-integral. The function $\epsilon(m)$ takes value $0$ if $m$ is integral, and takes value $1/2$ if $m$ is half-odd-integral.  The contribution of a chiral multiplet of charge $R$ coupled with a unit charge to the $\U(1)$ gauge multiplet such that the gauge field has a holonomy $z$ around the $\mathbb{S}^1$ and magnetic flux $m$ on the $\mathbb{S}^2$ is
\bes{
\CI^R_\chi(z; m; x) = (x^{1-R} z^{-1})^{|m|/2} \prod_{j=0}^\infty \frac{1-(-1)^m z^{-1} x^{|m|+2-R+2j}}{1-(-1)^m\, z\, x^{|m|-R+2j}}~.
}

It is worth pointing out that, for an integer-valued $m$, we have
\bes{ \label{qgauging}
\CI_{\CT_q}(w, n| f, m; x) &= \sum_{l' \in q\BZ} (w^\frac{1}{q}) ^{l'} \oint \frac{dz'}{2\pi i z'} z'^{\frac{n}{q}}~ \prod_{s=\pm1} \CI^{\frac{1}{2}}_\chi \left((z' f)^{s}; s (l' + m); x \right) \times  \\
& \qquad \qquad \qquad     \CI^{\frac{1}{2}}_\chi \left((z'^{-1} f)^{s}; s (-l' + m); x \right) \\
&= \frac{1}{q} \sum_{p=0}^{q-1} \CI_{\CT_1} \left(w^{\frac{1}{q}} e^{2\pi i \frac{p}{q}}, n/q | f, m; x \right)~,
}
where in the first equality we define $z'=z^q$ and $l' =q l$, and to establish the second equality we use the fact that, for an integer $\ell$, 
\bes{
\frac{1}{q} \sum_{p=0}^{q-1} \exp\left( 2\pi i \frac{\ell}{q} p\right) = \begin{cases} 1 & ~ \text{$q$ divides $\ell$} \\ 0 & ~ \text{$q$ does not divide $\ell$} \end{cases}~,
}
and so we have
\bes{
\sum_{l \in \BZ} \,\, \sum_{p=0}^{q-1} \frac{1}{q} w^{\frac{l}{q}}  \exp\left( 2\pi i \frac{l}{q} p\right) F(l)= \sum_{l' \in q\BZ} w^{\frac{l'}{q}} F(l')
}
for an arbitrary function of $F$.  The second equality of \eref{qgauging} is equivalent to the statement that one can obtain the $\CT_q$ theory from the $\CT_1$ theory by gauging a $\BZ_q$ subgroup of the $\U(1)_w$ symmetry associated with the fugacity $w$.  In 3d, gauging a zero-form symmetry leads to a new one-form symmetry.  Hence, the theory $\CT_q$ has a $\BZ_q$ one-form symmetry; in agreement with \cite{Bhardwaj:2022dyt}.  We will discuss the case of non-integer-valued $m$ in the subsequent subsections.

\subsection{The case of $q=1$}
For $q=1$, the theory $\CT_1$ is the $T(\SU(2))$ theory \cite{Gaiotto:2008ak}. The index of this theory as given by \eref{indTq} with the background magnetic fluxes $m=n=0$ is
\bes{ \label{indT1}
\scalebox{0.9}{$
\begin{split}
&\CI_{\CT_1}(w, n=0| f, m=0; x) \\
&= 1+ x\left[ (1+f^2+f^{-2}) + (1+w+w^{-1}) \right] + x^2 \left[ f^4+f^{-4} + w^2+w^{-2} -1\right] + \ldots\\
&= 1+ x \left[ \chi^{\mathfrak{su}(2)}_{[2]} (f) + \chi^{\mathfrak{su}(2)}_{[2]} (w) \right] \\
& \qquad + x^2 \left[\chi^{\mathfrak{su}(2)}_{[4]} (f) + \chi^{\mathfrak{su}(2)}_{[4]} (w) -\left(\chi^{\mathfrak{su}(2)}_{[2]} (f) + \chi^{\mathfrak{su}(2)}_{[2]} (w)+1 \right) \right] + \ldots~.
\end{split}$}
} 
Here the normalisation of the power of $w$ in \eref{indTq} is such that the object with magnetic flux $1$ corresponds to the fugacity $w$, and so the character of the adjoint representation $\chi^{\mathfrak{su}(2)}_{[2]} (w)$ written in terms of $w$ is $1+w+w^{-1}$.  However, the character of the adjoint representation $\chi^{\mathfrak{su}(2)}_{[2]} (f)$ written in terms of $f$ is $1+f^2+f^{-2}$.  As is well-known, 3d mirror symmetry maps the theory $\CT_1$ to itself \cite{Intriligator:1996ex} and interchanges $f^2$ with $w$ in this notation.

This theory has a global symmetry algebra $\mathfrak{su}(2)_f \oplus \mathfrak{su}(2)_w$, where the first and second $\mathfrak{su}(2)$ factors are referred to as the Higgs branch and Coulomb branch symmetries, respectively.  Since there is no odd power of the fugacity $f$ appearing in the index \eref{indT1}, there is no operator charged under the $\BZ_2$ centre of $\SU(2)_f$, and so the faithful Higgs branch symmetry is $\SU(2)_f/\BZ_2 \cong \SO(3)_f$. Similarly, since there is no half-odd-integral power of the fugacity $w$ appearing in the index \eref{indT1}, the faithful Coulomb branch symmetry is $\SU(2)_w/\BZ_2 \cong \SO(3)_w$.  Indeed, mirror symmetry interchanges the $\SO(3)_f$ and $\SO(3)_w$ symmetries.

In \cite{Gang:2018wek, Genolini:2022mpi, Bhardwaj:2022dyt}, it was pointed out that there is a mixed t'Hooft anomaly between $\SO(3)_f$ and $\SO(3)_w$, characterised by the 4d anomaly theory
\bes{ \label{mixedanomalyTSU2}
\exp \left( i \pi \int w_2^f \cup w_2^w \right)~.
}
where $w^{f,w}_2$ denotes the second Stiefel-Whitney class that is an obstruction to lifting the $\SO(3)_{f,w}$ bundles to the $\SU(2)_{f,w}$ bundles, and the integration is taken over a spin four-manifold whose boundary is the three-manifold in which the 3d theory in question lives.  In particular, as proposed in \cite[Section 7.2]{Bhardwaj:2022dyt} this mixed anomaly can be detected by examining the mixed gauge/zero-form monopole operators with fractional magnetic flux for both the gauge group and the Cartan subalgebra of the flavour symmetry group.  Our interpretation in terms of the index is to investigate such a mixed anomaly by setting the magnetic flux of the $\U(1)$ gauge group to be $\pm1/2$ and that of the Cartan subalgebra of $\SO(3)_f$ to be $1/2$.  This amounts to considering $\CF_{\CT_1}$ defined in \eref{indTq} with $l=\pm 1/2$, $m=1/2$ and $n=0$:
\bes{ \label{detectanomT1}
&\CF_{\CT_1} (w, n=0| f, m=1/2| l=\pm 1/2; x) \\
&= w^{\pm 1/2} \Big[ x^{1/2} +  x^{3/2} - (1+ f^2 + f^{-2}) x^{5/2}\\
& \qquad  +(3+ f^2 + f^{-2}) x^{7/2}+\ldots \Big]~,
}
where we emphasise that there is no half-odd-integral power of $f$ appearing.  Thus, such mixed gauge/zero-form monopole operators carry charge zero under the $\U(1)$ gauge symmetry, charge $\pm 1/2$ under the Cartan subalgebra of the $\mathfrak{su}(2)_w$ symmetry, and charge $0~(\mod~ 2)$ under the Cartan subalgebra of the $\mathfrak{su}(2)_f$ symmetry; in agreement with \cite[(7.85)]{Bhardwaj:2022dyt}.  This implies the mixed anomaly \eref{mixedanomalyTSU2}.  Moreover, one can use the index to check the consistency of other mixed gauge/zero-form monopole operators, \eg~ those with the fluxes $(1/3, 1/2)$ and $(1/4, 1/2)$ for the $\U(1)$ gauge group and the Cartan subalgebra of the $\SU(2)_f$, respectively, are not consistent because the integrand of the index contains fractional powers of the gauge fugacity, indicating that such fluxes are not properly quantised.

Another way to understand the anomaly is to gauge $\SO(3)_f$ or $\SO(3)_w$; let's say $\SO(3)_f$ for definiteness. This means that in the index we should integrated over $f$ and sum over $m\in\mathbb{Z}/2$. From the index result \eqref{detectanomT1} we see that the effect is of introducing half-integer spin representations of the topological symmetry, thus turning it into $\SU(2)_w$ as opposed to the original $\SO(3)_w$. This is the manifestation of the anomaly \eqref{mixedanomalyTSU2} at the level of the index, since this would be a gauge anomaly if we did not restrict to bundles for the topological symmetry that have $w_2^w=0$. We will come back to this point in Section \ref{sec:tsun}, where we will also generalise it to the $T(\SU(N))$ theory.

As a final remark, the $\CT_1$ theory can also be described as the $\SO(2)$ gauge theory with $1$ flavour of hypermultiplet in the vector representation, whose quiver can be depicted as
\bes{ \label{SO2w1vec}
\CT_1: \quad \SO(2) - [\USp(2)]
}
Of course, this theory is identical to $T(\SU(2))$ and so it maps to itself under 3d mirror symmetry.  However, if we view the $\CT_1$ theory as the $T(\USp(2))$ theory, its mirror theory is then $T(\SO(3))$ which can be described as \cite{Gaiotto:2008ak}
\bes{ \label{TSO3}
T(\SO(3)): \quad \O(1)-\USp(2)-[\SO(3)]
}
As explained in \cite{Gaiotto:2008ak, Assel:2018exy}, this description arises from gauging the $\BZ_2 \cong \O(1)$ symmetry in the $\USp(2)$ gauge theory with $\SO(4)$ flavour symmetry.  We will see that the description \eref{TSO3} provides a convenient way to come up with a mirror theory for the $\CT_2$ theory.  As a remark, in the description \eref{SO2w1vec} of the $\CT_1$ theory, only the $\BZ_2$ magnetic symmetry is manifest, and this indeed has a mixed anomaly with the $\SO(3)_f$ flavour symmetry.  This is characterised by the anomaly theory originated from \eref{mixedanomalyTSU2}, namely
\bes{\label{mixedanomalySO21vec}
\exp \left( i \pi \int w^f_2 \cup B_1^\CM \cup B_1^\CM \right)~,
}
where $B_1^\CM$ is the one-form background field for the magnetic symmetry.

\subsection{The case of $q=2$}
Let us first discuss some results for general $q$.  As we pointed out above, the theory $\CT_q$, namely the $\U(1)$ gauge theory with $2$ hypermultiplets of charge $q$, can be obtained from the theory $\CT_1$ by gauging a $\BZ_q$ subgroup of the $\SO(3)_w$ symmetry.  For $q>1$, the topological symmetry is $\U(1)_w$.  We find from \eref{indTq} that 
\bes{ \label{FTq}
&\CF_{\CT_q} (w, n=0| f, m=1/2| l=1/2q; x) \\
&= w^{\frac{1}{2q}} \Big[ x^{1/2} +  x^{3/2} - (1+ f^2 + f^{-2}) x^{5/2}\\
& \qquad  +(3+ f^2 + f^{-2}) x^{7/2}+\ldots \Big]~,
}
Following \cite{Bhardwaj:2022dyt}, the above result implies that the mixed gauge/zero-form monopole operator such that the flux for the $\U(1)$ gauge group is $\frac{1}{2q}$ and that for the Cartan subalgebra of $\SO(3)_f$ is $1/2$ carries charge $\frac{1}{2q}$ under the $\U(1)_w$ symmetry and charge $0~(\mod \,2)$ under the Cartan subalgebra of $\su(2)_f$.  This implies the mixed anomaly between the $\U(1)_w$ symmetry and the $\SO(3)_f$ symmetry for all $q$ characterised by the anomaly theory\footnote{From the result \eqref{FTq} we see that $c_1^w$ could in principle take value mod $2q$, but since $w_2^f$ is still valued mod 2 then also the entire anomaly is valued mod 2.}
\bes{\label{eq:anomq2}
\exp\left(i \pi \int w_2^f \cup c_1^w \,(\text{mod }2)\right)
}
where $c^w_1$ is the first Chern class associated with the $\U(1)_w$ topological symmetry of the $\CT_q$ theory.

Let us now focus on $q=2$.  Since  the $\CT_1$ theory can also be described by the $\SO(2)$ gauge theory as in \eref{SO2w1vec}, gauging its $\BZ_2$ zero-form magnetic symmetry leads to another description of the $\CT_2$ theory, namely the $\Spin(2)$ gauge theory with 1 flavour of hypermultiplet in the vector representation:
\bes{ \label{Spin21flv}
\CT_2: \quad \Spin(2)-[\USp(2)]
}
This theory has a {\it two-group symmetry} between the $\BZ_2$ one-form symmetry, arising from gauging the magnetic symmetry, and the flavour symmetry \cite[Section 7.4]{Bhardwaj:2022dyt}.  One can indeed see this as follows.  We can rewrite the anomaly theory \eref{mixedanomalySO21vec} for the $\CT_1$ theory as (see \cite[(2.19)]{Cordova:2017vab})
\bes{
\exp \left( i \pi \int w_2^f \cup \mathrm{Bock}(B_1^\CM) \right) = \exp \left( i \pi \int \mathrm{Bock}(w_2^f) \cup B_1^\CM \right)
}
The Postnikov class associated to the aforementioned two-group symmetry is then
\bes{
\delta B^\CM_2 = \mathrm{Bock}(w_2^f)
}
where $B^\CM_2$ is the two-form background field for the one-form symmetry arising from gauging the magnetic symmetry, and $\mathrm{Bock}$ is the Bockstein homomorphism associated with the short exact sequence:
\bes{
0 \, \longrightarrow \, \BZ_2 \, \longrightarrow \, \BZ_4 \, \longrightarrow \, \BZ_2 \, \longrightarrow \, 0~.
}
This indeed fits into the expectation that the $\Spin(4n+2)$ gauge theory with $N_f$ flavours in the vector representation has a two-group symmetry \cite{Benini:2018reh, Hsin:2020nts, Lee:2021crt, Apruzzi:2021mlh}.  We will also discuss this in more detail in Section \ref{sec:so}.  The index of the $\CT_2$ theory can be computed using \eref{indTq} with $q=2$:
\bes{
& \CI_{\CT_2}(w, n=0| f, m=0; x) \\
&= 1+ x \left[ 1+\chi^{\mathfrak{su}(2)}_{[2]}(f) \right] +x^2 \left[ \chi^{\mathfrak{su}(2)}_{[4]}(f) +w +w^{-1}  - \chi^{\mathfrak{su}(2)}_{[2]}(f)-1 \right] +\ldots
}
This indicates that the zero-form symmetry of the $\CT_2$ theory is $\SO(3)_f \times \U(1)_w$, where the faithful flavour symmetry is $\SO(3)_f$, not $\SU(2)_f$, since there is no $\su(2)_f$ representation with odd Dynkin label appearing in the index.  As pointed out in \cite[(7.129)]{Bhardwaj:2022dyt} and around \eref{FTq}, there is a mixed anomaly between the two-group symmetry and the $\U(1)_w$ symmetry.

It is worth pointing out that if we use the description \eref{Spin21flv} of the $\CT_2$ theory, instead of the $\U(1)$ gauge theory, the $\U(1)_w$ symmetry is not manifest and it should be regarded as emergent in the infrared.  The latter is, however, expected to be manifest as a flavour symmetry in the mirror theory, which we will shortly demonstrate below.  The mirror theory can be constructed as follows.  Recall that the $\CT_2$ theory can be obtained by gauging the $\BZ_2$ subgroup of the $\SO(3)_w$ topological symmetry of the $\CT_1$ theory. This should correspond to gauging the $\BZ_2 \cong \O(1)$ subgroup of the $\SO(3)$ flavour symmetry in the mirror theory \eref{TSO3} of the $\CT_1$ theory. We therefore propose that the mirror theory for $\CT_2$ is given by
\bes{ \label{mirrorT2}
\text{mirror of $\CT_2$}: \qquad  
\begin{tikzpicture}[baseline, font=\footnotesize]
\node[draw=none] (USp) at (0,0) {$\USp(2)$};
\node[draw=none] (O1a) at (-1.5,1) {$\O(1)$};
\node[draw=none] (O1b) at (-1.5,-1) {$\O(1)$};
\node[draw=none] (SO2f) at (2,0) {$[\SO(2)]$};
\draw (O1b)--(USp)--(O1a);
\draw (USp)--(SO2f);
\end{tikzpicture}
}
The $\SO(2)$ symmetry, which corresponds to the $\U(1)_w$ symmetry in the original $\CT_2$ theory, is indeed manifest in the mirror theory.  On the other hand, there is an emergent $\SO(3)$ symmetry in the mirror theory (see \cite{Gaiotto:2008ak}); this is mapped under mirror symmetry to the $\SO(3)_f$ flavour symmetry the original the $\CT_2$ theory.  Since we have gauged a $\BZ_2$ symmetry of the $T[\SO(3)]$ theory in order to construct \eref{mirrorT2}, the latter contains a $\BZ_2$ one-form symmetry, which is mapped to that of the original $\CT_2$ theory under mirror symmetry.  Since the $T[\SO(3)]$ theory is identical, as an SCFT, to the $T[\SU(2)]$ theory and hence the $\CT_1$ theory, there is a mixed anomaly between the $\SO(3)$ flavour symmetry and the $\SO(3)$ emergent magnetic symmetry in $T[\SO(3)]$.  As a result, gauging the $\BZ_2$ subgroup of the $\SO(3)$ flavour symmetry of $T[\SO(3)]$ leads to a non-trivial {\it two-group structure} between the new $\BZ_2$ one-form symmetry and the $\SO(3)$ emergent magnetic symmetry in \eref{mirrorT2}.  As expected, such a two-group symmetry in \eref{mirrorT2} maps to that of the $\CT_2$ theory under mirror symmetry.  We emphasise again that the origins of the $\BZ_2$ one-form symmetry and the $\SO(3)$ zero-form symmetry that participate in the two-group structure in the $\CT_2$ theory/its mirror \eref{mirrorT2} are interchanged under mirror symmetry in the following respective way: the $\BZ_2$ one-form symmetry arises from gauging of the $\BZ_2$ magnetic/flavour zero-form symmetry in the $\CT_1$ theory/its mirror \eqref{TSO3}, whereas the $\SO(3)$ symmetry is realised as the flavour/magnetic zero-form symmetry.

\section{Global symmetry group and anomalies of $T(\SU(N))$}
\label{sec:tsun}
One possible generalisation of our previous analysis for the $\U(1)$ gauge theory with two hypers of charge 1 is to the $T(\SU(N))$ theory of Gaiotto and Witten \cite{Gaiotto:2008ak}. This is a 3d $\mathcal{N}=4$ theory that can be described by the following quiver diagram:
\begin{equation}\label{TSUNquiver}
\begin{tikzpicture}[baseline,scale=0.85]
\tikzstyle{every node}=[font=\scriptsize]
\node[draw, circle] (p0) at (0,0) {\fontsize{14pt}{14pt}\selectfont $1$};
\node[draw, circle] (p1) at (3,0) {\fontsize{14pt}{14pt}\selectfont $2$};
\node (p2) at (5,0) {\fontsize{12pt}{12pt}\selectfont $\cdots$};
\node[draw, circle] (p3) at (7,0) {\fontsize{8pt}{8pt}\selectfont $N-1$};
\node[draw, rectangle] (p4) at (10,0) {\fontsize{14pt}{14pt}\selectfont $N$};

\draw (p0) to (p1);
\draw (p1) to (p2);
\draw (p2) to (p3);
\draw (p3) to (p4);
\end{tikzpicture}
\end{equation}
where each circle node denotes a unitary gauge group, the square node a flavour symmetry group and the lines hypermultiplets in bifundamental representations. Notice that $T(\SU(2))$ corresponds to the $\U(1)$ gauge theory with two hypers of charge 1. It is then natural to wonder how the discussion on the global form of the symmetry group and on the anomalies of $T(\SU(2))$ of the previous section generalises to $T(\SU(N))$. We will also discuss some consequences of these for theories that are obtained by gauging together various copies of $T(\SU(N))$, like the star-shaped quivers \cite{Benini:2010uu} that are the mirror duals to the 3d reduction of the 4d class $\mathcal{S}$ theories \cite{Gaiotto:2009we}.

The global symmetry algebra of the $T(\SU(N))$ theory is $\mathfrak{su}(N)_f\oplus\mathfrak{su}(N)_w$. The first piece $\mathfrak{su}(N)_f$ is the flavour symmetry acting on the hypermultiplets at the right end of the quiver, which is special unitary and not unitary since the $\mathfrak{u}(1)$ baryonic symmetry can be reabsorbed with a gauge transformation. The second factor $\mathfrak{su}(N)_w$ is the topological symmetry that acts on monopole operators. This is enhanced in the quiver description \eqref{TSUNquiver} where only the Cartan subalgebra is manifest. Such a symmetry enhancement is due to the fact that all the gauge nodes are \textit{balanced}, that is they have a number of flavours which is twice the number of colours, which implies that there are monopole operators with dimension 1 that supplement the required extra moment maps \cite{Gaiotto:2008ak}.

The actual global symmetry group is $\PSU(N)_f\times \PSU(N)_w$. This is easy to see for the flavour symmetry, since one can use a gauge transformation to also reabsorb a transformation of the centre of $\SU(N)$ so that the actual flavour symmetry group is $\U(N)/\U(1)=\PSU(N)_f$. One way to argue that also the topological symmetry is $\PSU(N)_w$ is by resorting to mirror symmetry, under which the $T(\SU(N))$ theory is self-dual as it can be easily understood from the Hanany--Witten brane set-up of the theory \cite{Hanany:1996ie}. Even though the $T(\SU(N))$ theory is simply dual to itself, there is still a non-trivial map on the flavour and topological symmetries which are exchanged by the mirror duality. Hence, the two global symmetry groups should be identical. Another way to understand what is the global symmetry group is using the index. As we will show later for low $N$, computing the index one indeed finds representations of $\mathfrak{su}(N)_f$ and $\mathfrak{su}(N)_w$ that are uncharged under the $\mathbb{Z}_N$ centre. Notice that this generalises the case of $T(\SU(2))$, for which we saw that the global symmetry group is $\SO(3)_f\times \SO(3)_w$.

The $\PSU(N)$ group admits bundles with non-trivial second Stiefel--Whitney class, which measures the obstruction to lifting them to $\SU(N)$ bundles. Using the second Stiefel--Whitney classes $w_2^{f,w}$ of $\PSU(N)_{f,w}$ one can in principle write a discrete anomaly between them. We propose that in the $T(\SU(N))$ theory there is indeed such an anomaly, which generalises the one in \eqref{mixedanomalyTSU2} for $N=2$
\bes{ \label{mixedanomalyTSUN}
\exp \left( \frac{2i \pi}{N} \int w_2^f \cup w_2^w \right)~.
}

Before providing evidence for these claims using the superconformal index, let us discuss some of their consequences. The first one is that if we gauge one of the two symmetries, say the flavour symmetry, as $\PSU(N)_f$ rather than $\SU(N)_f$ then the other symmetry becomes $\SU(N)_w$. This is because in the gauging we sum over bundles with non-trivial $w_2^f$, so we necessarily have to restrict to bundles with trivial $w_2^w$ to avoid the anomaly \eqref{mixedanomalyTSUN} which has become a gauge anomaly after the gauging. At the level of the index, as we will see, one can observe that considering a suitable fractional background flux for the $\PSU(N)_f$ symmetry introduces states that transform in representations which are charged under the centre of $\SU(N)_w$. Because of this, if we want to gauge both symmetries of $T(\SU(N))$, at most one of them can be gauged as $\PSU(N)$ and not both. 

In a similar manner, one can understand the precise global structure of the symmetry group of the 4d theories of class $\mathcal{S}$ from their 3d mirror star-shaped quivers. Consider for example the $T_N$ theory, which is realised in class $\mathcal{S}$ by compactifying the 6d $(2,0)$ theory of type $A_{N-1}$ on a sphere with three regular maximal punctures. The associated 3d mirror is a star-shaped quiver with three $T(\SU(N))$ legs that are glued together by gauging a common $\PSU(N)$ symmetry. Suppose that we gauge a diagonal combination of the flavour symmetry of each $T(\SU(N))$. Then, due to \eqref{mixedanomalyTSUN} we have a mixed gauge anomaly between the middle $\PSU(N)$ gauge group of the star-shaped quiver and the three remaining topological symmetries of the form $\exp \left( \frac{2i \pi}{N} \int w_2 \cup \left(w_2^{w,1}+w_2^{w,2}+w_2^{w,3}\right) \right)$, where $w_2$ is for the middle gauge group and $w_2^{w,i}$ with $i=1,2,3$ are for the three topological symmetries. This means that the global symmetry group of the theory should be such that the combination $w_2^{w,1}+w_2^{w,2}+w_2^{w,3}$ is trivial, which implies that only a diagonal combination of the three $\mathbb{Z}_N$ centres of the topological symmetries is allowed to act non-trivially on the spectrum of the theory. We then recover the known result that the global symmetry group of $T_N$ is $\left(\SU(N)\times \SU(N)\times \SU(N)\right)/\left(\mathbb{Z}_N\times\mathbb{Z}_N\right)$ \cite{Gadde:2011ik,Bhardwaj:2021ojs}. For example, for $N=2$ we have the $T_2$ theory whose 3d mirror is the diagonal $\SO(3)$ gauging of three copies of the $T(\SU(2))$ theory\footnote{The fact that the middle node of the star-shaped quiver should be taken to be $\SU(2)/\BZ_2 \cong \SO(3)$ instead of $\SU(2)$ in order to obtain the correct mirror of the theory of eight free half-hypermultiplets and not a $\mathbb{Z}_2$ discrete gauging thereof was emphasised in \cite{Cremonesi:2014vla, Razamat:2014pta}.} and which coincides with the theory of eight free half-hypermultiplets in the $[1;1;1]$ representation of the global symmetry $\mathfrak{su}(2)^3$ which is charged under the diagonal $\mathbb{Z}_2$ centre. 

This reasoning can also be extended to other class $\mathcal{S}$ theories. For example, one may also take four copies of the $T(\SU(2))$ theory and gauge the diagonal $\SO(3)$ subgroup of $\SO(3)^4_f$.  The resulting theory is the mirror theory of the $\SU(2)$ gauge theory with $4$ flavours of fundamental hypermultiplets, which is the class $\mathcal{S}$ theory of type $A_1$ on a sphere with four regular punctures.  In this case the $\mathfrak{su}(2)^4_w$ symmetry of the star-shaped quiver gets enhanced to the $\mathfrak{so}(8)$ flavour symmetry of the latter theory in the infrared.  There is a moment map operator (the mesons) in the adjoint representation of $\mathfrak{so}(8)$, where upon applying the branching rule, it contains the representation $[1;1;1;1]$ of $\mathfrak{su}(2)^4_w$, which is charged under the diagonal $\mathbb{Z}_2$ centre. One can also check that all the representations appearing in the spectrum are only charged under the  diagonal $\mathbb{Z}_2$ centre of $\mathfrak{su}(2)^4_w$. This fact can again be understood as a consequence of the anomaly \eqref{mixedanomalyTSUN}.

In order to argue that the global symmetry group of $T(\SU(N))$ is $\PSU(N)_f\times \PSU(N)_w$ and that there is the anomaly \eqref{mixedanomalyTSUN} we will use the superconformal index. The index of $T(\SU(N))$ can be expressed recursively as follows:
\bes{ \label{indTSUN}
\CI_{T(\SU(N))}(\vec{w}, \vec{n}| \vec{f}, \vec{m}; x) &= \frac{1}{(N-1)!}\sum_{\vec{l}\in\mathbb{Z}^N+\vec{\epsilon}(\vec{m})}\prod_{a=1}^{N-1} w_{N-1}^{l_a} \oint \prod_{a=1}^{N-1}\frac{dz_a}{2\pi i z_a} z_a^{n_{N-1}}\times  \\
&\prod_{i=1}^{N}\prod_{a=1}^{N-1}\prod_{s=\pm1} \CI^{\frac{1}{2}}_\chi \left((z_a f_i)^{s}; s (l_a + m_i); x \right) \times  \\
&\CI_{T(\SU(N-1))}(\{w_1,\cdots,w_{N-2}\}, \{n_1,\cdots,w_{N-2}\}| \\
&\qquad\qquad\qquad\{z_1,\cdots,z_{N-1}\}, \{l_1,\cdots,n_{N-1}\}; x)~, 
}
where $\vec{w}, \vec{n}$ and $\vec{f}, \vec{m}$ are the fugacities and background magnetic fluxes for the topological and the flavour symmetry\footnote{As usual, we parametrize the flavour symmetry fugacities by $f_i$ and the fluxes by $m_i$ with $i=1,\cdots,N$ with the constraints $\prod_{i=1}^Nf_i=1$ and $\sum_{i=1}^Nm_i=0$.} respectively and the sum over the gauge fluxes $\vec{l}$ depends on the fractional part of the background fluxes for the global symmetries which we encode in $\vec{\epsilon}(\vec{m})$.

Let us consider the case in which all the background fluxes are turned off, that is $\vec{n}=\vec{m}=(0,\cdots,0)$. In this case $\vec{\epsilon}(\vec{m})$ is trivial and so all the gauge fluxes should be taken to be integers. Computing the index perturbatively to low orders in the fugacity $x$ we find for $N=3,4$ (the result for $N=2$ can be found in \eqref{indT1})\footnote{The index of the $T(\SU(N))$ theory was computed also in \cite{Beratto:2020qyk}, see eq.~(2.1), with the further refinement of a fugacity for the axial symmetry which is the commutant of the $\mathcal{N}=2$ R-symmetry inside the $\mathcal{N}=4$ R-symmetry. Moreover, it was pointed out that the $-1$ at order $x^2$ corresponds to the extra SUSY current from the point of view of the $\mathcal{N}=3$ index, see also \cite{Evtikhiev:2017heo,Gang:2018huc,Garozzo:2019ejm,Gang:2021hrd}.}
\bes{ \label{indTSUNlowN}
N=3:\quad &1+\left(\chi_{[1,1]}(\vec{f})+\chi_{[1,1]}(\vec{w})\right)x+\\
&\quad\left(\chi_{[2,2]}(\vec{f})+\chi_{[2,2]}(\vec{w})+\chi_{[1,1]}(\vec{f})\chi_{[1,1]}(\vec{w})-1\right)x^2+\cdots~,\\
N=4:\quad &1+\left(\chi_{[1,0,1]}(\vec{f})+\chi_{[1,0,1]}(\vec{w})\right)x+\\
&\quad\left(\chi_{[2,0,2]}(\vec{f})+\chi_{[2,0,2]}(\vec{w})+\chi_{[1,0,1]}(\vec{f})\chi_{[1,0,1]}(\vec{w})+\right.\\
&\quad\left.\chi_{[0,2,0]}(\vec{f})+\chi_{[0,2,0]}(\vec{w})-1\right)x^2+\cdots~,
}
where $\chi_{[k_1,\cdots,k_N]}(\vec{f})$ denotes the character of the representation of $\mathfrak{su}(N)_f$ with Dynkin label $[k_1,\cdots,k_N]$ and similarly for $\chi_{[k_1,\cdots,k_N]}(\vec{w})$. We can see that the only representations that appear are those that are uncharged under the $\mathbb{Z}_N$ centre,\footnote{One can check that this is true also to higher orders in the expension in $x$ and for higher $N$.} implying that the global symmetry group is indeed $\PSU(N)_f\times \PSU(N)_w$.

We can then compute the index with a non-trivial value of the background fluxes for this global symmetry which are fluxes of $\PSU(N)$ that are not fluxes of $\SU(N)$ or more generally of $\SU(N)/\mathbb{Z}_k$ for any $k$ that divides $N$ and which is not $N$ itself. For example, we can consider the $\PSU(N)_f$ flux $\vec{m}=\left(\frac{1}{N},\cdots,\frac{1}{N},-\frac{N-1}{N}\right)$ while we take the $\PSU(N)_w$ flux to be trivial. In this case, the gauge fluxes should be taken to be in $\mathbb{Z}-\frac{1}{N}$ so to have a correct quantization of the gauge charges in the monopole background. The first non-trivial contribution to the expansion in $x$ of the index for low $N$ is
\bes{ \label{indTSUNlowNrac}
N=3:\quad &\chi_{[1,0]}(\vec{w})x+\cdots~,\\
N=4:\quad &\chi_{[1,0,0]}(\vec{w})x^{\frac{3}{2}}+\cdots~.
}
Looking also at \eqref{detectanomT1} we can expect that the contribution for generic $N$ is $\chi_{[1,0,\cdots,0]}(\vec{w})x^{\frac{N-1}{2}}$. The Dynkin label $[1,0,\cdots,0]$ corresponds to the fundamental representation of $\SU(N)$ which has charge $1$ under the $\mathbb{Z}_N$ centre. This implies the anomaly \eqref{mixedanomalyTSUN} in the $T(\SU(N))$ theory. In particular, gauging the flavour symmetry as a $\PSU(N)_f$ symmetry would turn the topological symmetry into an $\SU(N)_w$ symmetry, since in the index we should sum over fluxes of the form $\vec{m}=\left(\frac{1}{N},\cdots,\frac{1}{N},-\frac{N-1}{N}\right)$ which introduces states in the representations of $\SU(N)_w$ that are not representations of $\PSU(N)_w$ or any other $\SU(N)_w/\mathbb{Z}_k$.

\section{$\so(2N)_k$ gauge algebra and $N_f$ hypermultiplets}
\label{sec:so}
Let us now extend our results of Section \ref{sec:u1} to the 3d $\CN=3$ gauge theory with $\so(2N)_k$ gauge algebra and with $N_f$ flavours of hypermultiplets in the vector representation.  Let us first summarise the main results for the $\SO(2N)_k$ gauge group and then provide the evidence and reasons later. 
\bi
\item For $k$ odd, the dressed monopoles that are gauge invariant involve an odd number of chiral fields, and so the global form of the flavour symmetry is $\USp(2N_f)$, not $\USp(2N_f)/\BZ_2$.  There is no discrete anomaly involving the 2nd Stiefel-Whitney class associated to the flavour symmetry bundle.
\item For $k$ divisible by $4$, the global form of the flavour symmetry is $\USp(2N_f)/\BZ_2$, and the anomaly theories are
\bes{ \label{anomk4}
\begin{tabular}{|c|c|c|}
\hline
$N$ & $N_f$ & Anomaly theory \\
\hline
{\blue even} & {\blue even} & {\blue $\exp\left(i \pi \int B^\CM_1 \cup B^\CC_1 \cup w_2^f  \right)$} \\
even & odd & $\exp\left(i \pi \int B^\CC_1 \cup (B^\CM_1 + B^\CC_1) \cup w_2^f  \right)$ \\
{\blue odd} & {\blue even} & {\blue $\exp\left(i \pi \int B^\CM_1 \cup (B^\CM_1 + B^\CC_1) \cup w_2^f  \right)$} \\
odd & odd & $\exp\Big(i \pi \int B^\CM_1 \cup B^\CM_1 \cup w_2^f  + B^\CC_1 \cup B^\CM_1 \cup w_2^f$ \\ 
 & & $+  B^\CC_1 \cup B^\CC_1 \cup w_2^f \Big)$ \\
\hline
\end{tabular}
}
where $B_1^{\CM}$ ({\it resp.} $B_1^{\CC}$) is the one-cocycle that is the background field for the $\BZ_2$ zero-form magnetic symmetry $\CM$ ({\it resp.} the $\BZ_2$ zero-form charge conjugation symmetry $\CC$), and $w_2^f$ is the 2nd Stiefel-Whitney class that obstructs the lift of the $\USp(2N_f)/\BZ_2$ bundles to the $\USp(2N_f)$ bundles. The two rows highlighted in blue are exchanged under the duality, which will be discussed below, and each row in black is mapped to itself under the duality.
\item  We will discuss the case of $k \equiv 2 \, (\mod \, 4)$ and its subtleties in Section \ref{sec:level4Kp2}. 
\ei

In the following discussion we will focus on the case in which $k = 4K$ is divisible by $4$ (\ie, $K \in \BZ$). Gauging the magnetic symmetry in the above $\SO(2N)_{4K}$ gauge theory by making $B_1^\CM$ dynamical, we obtain the $\Spin(2N)_{4K}$ gauge theory with $N_f$ hypermultiplets in the vector representation, whose two-group symmetries are given by
\bes{ \label{twogroupSpin2N}
&\qquad\qquad \text{$\Spin(2N)_{4K}$ + $N_f$ flavours} \\
&\begin{tabular}{|c|c|c|}
\hline
$N$ & $N_f$ & Postnikov class \\
\hline
even & even & $\delta B^\CM_2 =  B^\CC_1 \cup w_2^f $\\
even & odd & $ \delta B^\CM_2 =  B^\CC_1 \cup w_2^f$ \\
odd & even & $ \delta B^\CM_2 =  \mathrm{Bock}(w_2^f) + B^\CC_1  \cup w_2^f$ \\
odd & odd & $ \delta B^\CM_2 =   \mathrm{Bock}(w_2^f) + B^\CC_1 \cup w_2^f$ \\
\hline
\end{tabular}
}
where $B^\CM_2$ is the two-form background field associated with the one-form symmetry arising from gauging the magnetic symmetry. In the above, we followed \cite[(2.19)]{Cordova:2017vab} and rewrote the relevant terms in the third and fourth lines in \eref{anomk4} as
\bes{
\pi \int  w_2^f \cup B_1^\CM \cup B_1^\CM   = \pi \int  w_2^f  \cup \mathrm{Bock}( B_1^\CM)  = \pi \int  \mathrm{Bock}(w_2^f) \cup B_1^\CM ~,
}
where $\mathrm{Bock}$ is the Bockstein homomorphism associated with the short exact sequence $0 \rightarrow \BZ_2 \rightarrow \BZ_4 \rightarrow \BZ_2 \rightarrow 0$.  Similarly to the discussion in \cite[(2.20)]{Cordova:2017vab}, upon making $B_1^\CM$ dynamical, we have $\delta B^\CM_2 =  \mathrm{Bock}(w_2^f)$, as required.

On the other hand, gauging the magnetic symmetry in the above $\SO(2N)_{4K}$ gauge theory by making $B_1^\CC$ dynamical, we obtain the $\O(2N)^+_{4K}$ gauge theories with $N_f$ hypermultiplets in the vector representation, whose two-group symmetries are given by
\bes{ \label{twogroupO2N}
&\qquad\qquad \text{$\O(2N)^+_{4K}$ + $N_f$ flavours} \\
&
\begin{tabular}{|c|c|c|}
\hline
$N$ & $N_f$ & Postnikov class \\
\hline
{\blue even} & {\blue even} & {\blue $\delta B^\CC_2 =  B^\CM_1 \cup w_2^f $} \\
even & odd & $ \delta B^\CC_2 =  \mathrm{Bock}(w_2^f) + B^\CM_1 \cup w_2^f$ \\
{\blue odd} & {\blue even} & {\blue $ \delta B^\CC_2 =  B^\CM_1  \cup w_2^f$} \\
odd & odd & $ \delta B^\CC_2 =  \mathrm{Bock}(w_2^f) + B^\CM_1 \cup w_2^f$ \\
\hline
\end{tabular}
}
Again, the two rows highlighted in blue are exchanged under the duality, whereas each row in black is mapped to itself under the duality.

For $k=0$, the presence of the mixed anomaly in the $\SO(4n+2)_0$ gauge theory and the two-group in the $\Spin(4n+2)_0$ gauge theory meets the usual expectation, as pointed out in \cite{Lee:2021crt, Apruzzi:2021mlh} (see also \cite{Benini:2018reh, Hsin:2020nts}). For $k \geq 0$, the above statements regarding the $\SO(2N)_{4K}$ gauge theories can be seen from the indices which can be computed in the following way. For the fugacity of the charge conjugation symmetry $\chi=+1$ ({\it resp.} $\chi=-1$), all gauge magnetic fluxes are set to be $\pm 1/2$ ({\it resp.} except the last one $m_{N}$ is set to zero and the gauge fugacities $z_N$ and $z^{-1}_N$ are set to $1$ and $-1$ respectively) and all flavour background magnetic fluxes are fixed to be $1/2$ (we refer the reader to Appendix \ref{app:index} for a summary of the relevant facts about the 3d superconformal index).  We provide some examples for the $\SO(2N)_k$ gauge theory with $N_f$ flavours of hypermultiplets in the vector representation in the table below.\footnote{These results are up to possible overall minus signs.}  
\bes{
\scalebox{0.8}{
\begin{tabular}{|c|c|c|c|c|c|}
\hline
$N$ & $N_f$ & $k$ & $\chi$& index \\
\hline
$2$  & &  $0$ & $+1$ & $\frac{\zeta}{2} \left[ x^{N_f-1} + x^{N_f}  \left(N_f+ \sum_{1\leq i\neq j \leq N_f} f_i f_j^{-1} \right) +\ldots \right] $ \\
       & $\begin{Bmatrix} \text{even} \\ \text{odd} \end{Bmatrix}$ &         & $-1$ &  $\begin{Bmatrix} 1 \\ i \end{Bmatrix} \times \frac{\zeta^{\frac{1}{2}}}{2} \left[ x^{N_f-1} + x^{N_f}  \left(N_f+ \sum_{1\leq i\neq  j \leq N_f} f_i f_j^{-1} \right) +\ldots \right] $ \\
\hline
$2$  & &  $4$ & $+1$ & $\frac{\zeta}{2} \left[  x^{N_f+1}  \left( \text{polynomials in $(f_1, \ldots, f_{N_f})$ of degrees 2, 4} \right) +\ldots \right] $ \\
       & $\begin{Bmatrix} \text{even} \\ \text{odd} \end{Bmatrix}$ &         & $-1$ &  $\begin{Bmatrix} 1 \\ i \end{Bmatrix} \times \frac{\zeta^{\frac{1}{2}}}{2} \left[ x^{N_f}  \left( \text{polynomials in $(f_1, \ldots, f_{N_f})$ of degrees 0, 2} \right) +\ldots \right] $ \\
\hline
\hline
$3$ & & $0$ & $+1$ & $\frac{\zeta^{\frac{3}{2}}}{4} \left[ x^{\frac{3N_f-6}{2}}+x^{\frac{3N_f-4}{2}} \left(N_f+ \sum_{1\leq i\neq j \leq N_f} f_i f_j^{-1} \right)+\ldots \right] $ \\ 
 &   $\begin{Bmatrix} \text{even} \\ \text{odd} \end{Bmatrix}$ & &   $-1$ & $\begin{Bmatrix} 1 \\ i \end{Bmatrix} \times \frac{\zeta}{4} \left[ x^{\frac{3N_f-6}{2}}+ x^{\frac{3N_f-4}{2}} \left(N_f+ \sum_{1\leq i\neq j \leq N_f} f_i f_j^{-1} \right)+\ldots \right] $ \\
\hline 
$3$ & & $4$ & $+1$ & $\frac{\zeta^{\frac{3}{2}}}{4} \left[ x^{\frac{3N_f}{2}}\left( \text{polynomials in $(f_1, \ldots, f_{N_f})$ of degrees 0, 2, 4, 6} \right)+\ldots \right] $ \\ 
 &   $\begin{Bmatrix} \text{even} \\ \text{odd} \end{Bmatrix}$ & &   $-1$ & $\begin{Bmatrix} 1 \\ i \end{Bmatrix} \times \frac{\zeta}{4} \left[ x^{\frac{3N_f-2}{2}} \left( \text{polynomials in $(f_1, \ldots, f_{N_f})$ of degrees 2, 4} \right)+\ldots \right] $ \\
\hline 
\end{tabular}
}
}

These results indicate the presence of the mixed anomalies given by Table \eref{anomk4}.  This can be explained via examples as follows.  Let us first consider an example for $N$ even and $N_f$ even. We can set $\chi=1$ (\ie~ turn off $B_1^\CC$) and sum over $\zeta=\pm 1$ without getting any imaginary number in the coefficients; this indicates that, if $B_1^\CC$ is turned off, there is no anomaly that obstructs gauging the magnetic symmetry and the flavour symmetry $\USp(2N_f)/\mathbb{Z}_2$ simultaneously, \ie~ no mixed anomaly involving $B_1^\CM$.  On the other hand, if we set $\chi =-1$ (\ie~ turn on $B_1^\CC$), summing over $\zeta=\pm 1$ yields an imaginary coefficient. This means that there is an obstruction in gauging the magnetic symmetry if the background field for the charge conjugation symmetry and a non-trivial $\USp(2N_f)/\mathbb{Z}_2$ bundle are turned on, \ie~ there is a mixed anomaly $\exp \left(i \pi \int B_1^\CM \cup B_1^\CC \cup w_2^f  \right)$.  Let us now consider an example for $N$ even and $N_f$ odd.  If we set $\zeta=1$ (\ie~ turn off $B_1^\CM$), summing over $\chi = \pm 1$ yields an imaginary coefficient, but if we set $\zeta = -1$ (\ie~ turn on $B_1^\CM$), we can sum over $\chi=\pm 1$ without getting an imaginary coefficient. This implies that, if we turn off $B_1^\CM$, we have the anomaly $\exp \left(i \pi \int B_1^\CC \cup B_1^\CC \cup w_2^f  \right)$.  However, if we turn on $B_1^\CM$, this anomaly gets cancelled by an additional factor $\exp \left(i \pi \int B_1^\CC \cup B_1^\CM \cup w_2^f  \right)$.  In terms of the index, each of these anomaly factors contributes an imaginary number, and multiplying two imaginary numbers yields a real number.  In conclusion, the anomaly theory is $\exp \left(i \pi \int B_1^\CC \cup (B_1^\CM+B_1^\CC) \cup w_2^f  \right)$ as required.  Note that for $k$ divisible by 4, the flavour fugacities $f_i$ (with $i=1, \cdots, N_f$) always appear as polynomials of even degrees, and so there is no anomaly involving $w_2^f \cup w_2^f$.    This analysis can be carried out generally.  For convenience, we summarise the correspondence between the anomaly theory and its contribution to the index, where the gauge magnetic fluxes and flavour background magnetic fluxes are fixed to be $1/2$ as follow\bes{ \label{anomindex}
\scalebox{0.85}{
\begin{tabular}{|c|c|}
\hline
Anomaly theory & Feature of the index \\
\hline
$\exp \left( i\pi  \int B^\CM_1 \cup B^\CM_1 \cup w_2^f \right)$ & $\zeta^{\pm \frac{1}{2}}$  for the indices for both $\chi = \pm 1$, \ie~ only \\
&  $(\zeta = -1, \chi = \pm 1)$ give imaginary coefficients.\\ 
 \hline
$\exp \left( i \pi  \int B^\CC_1 \cup B^\CC_1 \cup w_2^f \right)$ & $\chi^{\pm \frac{1}{2}}$  for the indices for both $\zeta = \pm 1$, \ie~ only \\
 & $(\zeta = \pm 1, \chi = - 1)$ give imaginary coefficients\\ 
 \hline
$\exp \left( i \pi  \int B^\CM_1 \cup B^\CC_1 \cup w_2^f \right)$ & $\zeta$ for $\chi = + 1$ and $\zeta^{\pm \frac{1}{2}}$ for $\chi = - 1$, \ie~ only \\
 & $(\zeta = -1, \chi = - 1)$ gives imaginary coefficients\\ 
\hline
$\exp \left( i \pi \int B^\CM_1 \cup (B^\CM_1+ B^\CC_1) \cup w_2^f\right)$  & $\zeta^{\pm \frac{1}{2}}$ for $\chi = + 1$ and $\zeta$ for $\chi = - 1$, \ie~ only \\
&  $(\zeta = -1, \chi = + 1)$ gives imaginary coefficients.\\
 \hline
$\exp \left( i \pi \int B^\CC_1 \cup (B^\CC_1+ B^\CM_1) \cup w_2^f \right)$ & $\chi^{\pm \frac{1}{2}}$ for $\zeta= + 1$ and $\chi$ for $\zeta = - 1$, \ie~ only \\
 & $(\zeta = + 1, \chi = -1)$ gives imaginary coefficients\\
\hline
$\exp\Big(i \pi \int B^\CM_1 \cup B^\CM_1 \cup w_2^f  $ & $\zeta^{\pm \frac{1}{2}}$ for $\chi= + 1$ and $(-\zeta)^{\pm \frac{1}{2}}$ for $\chi = - 1$, \ie~ only $(\zeta, \chi) =$ \\
$\qquad +  B^\CC_1 \cup B^\CC_1 \cup w_2^f \Big)$ &  $(+1,-1)$, $(- 1,+1)$ give imaginary coefficients\\
 \hline
$\exp\Big(i \pi \int B^\CM_1 \cup B^\CM_1 \cup w_2^f  $ & $\zeta^{\pm \frac{1}{2}}$ for $\chi= + 1$ and $i \zeta$ for $\chi = - 1$, \ie~ only $(\zeta, \chi) =$ \\
$+ B^\CC_1 \cup B^\CM_1 \cup w_2^f +  B^\CC_1 \cup B^\CC_1 \cup w_2^f \Big)$ &  $(-1,+1)$, $(\pm 1,-1)$ give imaginary coefficients\\
 \hline
\end{tabular}
}
}

We will provide a non-trivial test of these results in the next subsection using the duality.

\subsection{Compatibility with the duality}

Let us provide another indirect argument to support the above claims.  We utilise the duality between the following 3d $\CN=2$ gauge theories \cite[Section 5.3]{Aharony:2013kma}: (a) the $\SO(n_c)_k$ ({\it respectively} $\O(n_c)^+_k$ and $\Spin(n_c)_k$) gauge theory with $n_f$ flavours of chiral multiplets in the vector representation and zero superpotential, and (b) the $\SO(n_c')_{-k}$ ({\it resp.} $\O(n_c')^+_{-k}$ and $\O(n_c')^-_{-k}$) gauge theory, $n'_c=  n_f+|k|-n_c+2$, with $n_f$ flavours of chiral multiplets $q$ in the vector representation, a collection of $n_f(n_f+1)/2$ gauge singlets $M$, and superpotential $W=M q q $.  Using the same argument as in \cite{Kapustin:2011gh}, we can establish a duality between the following 3d $\CN=3$ gauge theories:\footnote{In the 3d $\mathcal{N}=3$ theory the CS coupling gives a mass to the $\mathcal{N}=2$ adjoint chiral multiplet inside the $\mathcal{N}=3$ vector multiplet. Integrating it out we obtain an effective $\mathcal{N}=2$ theory with a quartic superpotential for the massless chiral fields in the vector representation. The $\mathcal{N}=3$ duality can then be understood as a consequence of the $\mathcal{N}=2$ duality of \cite{Aharony:2013kma} deformed by such quartic coupling.}
\bes{ \label{dualityN3}
& \text{(1) the $\SO(N_c)_k$ ({\it resp.} $\O(N_c)^+_k$ and $\Spin(N_c)_k$) gauge theory} \\ 
& \qquad \text{ with $N_f$ hypermultiplets in the vector representation, and}\\
& \text{(2) the $\SO(N_c')_{-k}$ ({\it resp.} $\O(N_c')^+_{-k}$ and $\O(N'_c)^{-}_{-k}$) gauge theory,}  \\
& \qquad \text{$N'_c= 2N_f +|k| -N_c +2$, with $N_f$ hypermultiplets} \\
& \qquad \text{in the vector representation.}
}

Let us focus first on the special orthogonal $\SO$ gauge group. As pointed out around \cite[(6.12)]{Aharony:2013kma}, in order to match the index of the two theories, one needs to include to the index of theory (2) the contact term $\prod_{i=1}^{N_f} f_i^{k n^{(i)}_f}$, where $f_i$ are the flavour fugacities and $n^{(i)}_f$ are the background magnetic fluxes for the flavour symmetry.  If we denote by $(\zeta, \chi)$ and $(\zeta', \chi')$ the fugacities for $(\text{magnetic}, \text{charge conjugation})$ symmetries of the theories (1) and (2) respectively, then we have a fugacity map
\bes{ \label{zetachimap}
\zeta' = \zeta~, \quad \chi' = \zeta \chi~, \quad \text{or equivalently} \quad  \zeta = \zeta'~, \quad \chi = \zeta' \chi'~.
}
It can easily be seen that \eref{anomindex} is consistent with the duality, where rows $(1, 2, 3, 5, 7)$ get exchanged with rows $(1, 6, 4, 5, 7)$, respectively.

Suppose that we take $N_c$ to be an even number, say $N_c=2N$. Then, the number of colours $N_c'  = 2N_f +|k| - 2N +2$ in theory (2) is even if $k$ is even while it is odd if $k$ is odd.  In particular, the duality \eref{dualityN3} exchanges the $\SO(\text{even})_{\text{odd}}$ gauge theory with $\SO(\text{odd})_{\text{odd}}$ gauge theory.\footnote{A benefit of dealing with the latter is that its index can be easily computed for both $\chi  = +1$ and $\chi=-1$ at the same time, without having to do two separate computations as for the $\SO(\text{even})_{\text{odd}}$ gauge theory \cite{Aharony:2013kma} (see also Appendix \ref{app:index}).} The latter theory has a flavour symmetry $\USp(2N_f)$, rather than $\USp(2N_f)/\mathbb{Z}_2$.  This is due to the fact that $\SO(\text{odd})$ has a trivial centre and so the $\BZ_2$ centre of the $\USp(2N_f)$ flavour symmetry cannot be reabsorbed into that of the gauge group $\SO(\text{odd})$; therefore we cannot have a discrete anomaly for this theory. This is compatible with our findings for the $\SO(\text{even})_{\text{odd}}$ theory.

Let us now explore the case of $\SO(N_c)_{k}$ gauge theory with $N_f$ flavours, where $N_c=2N$ is even and $k = 2\kappa$ is even. The number of colours of the dual theory is $N_c'  = 2(N_f + |\kappa |- N) +2$, which can be both $0\,(\text{mod}\,4)$ or $2\,(\text{mod}\,4)$ depending on the level and the number of flavours.  In particular, we see that $N_c/2$ and $N_c'/2$ have the same partity, \ie ~ both being odd or even, if and only if $N_f + |\kappa|$ is odd; otherwise they have the opposite parity. Let us further assume that $k$ is divisible by $4$, \ie~ $\kappa$ is even.\footnote{The case of $k \equiv 2 \, (\mod\, 4)$, \ie~ $\kappa$ is odd, will be considered in Section \ref{sec:level4Kp2}.}  As a consequence, if $N_f$ is odd, then $N_c/2$ and $N_c'/2$ have the same parity, and each anomaly denoted in black in \eref{anomk4} is mapped into itself, in accordance with the duality \eref{zetachimap} where $\CC \leftrightarrow \CC \CM$ and $\CM \leftrightarrow \CM$.  On the other hand, if $N_f$ is even, then $N_c/2$ and $N_c'/2$ have opposite parity, and the two anomalies denoted in blue in \eref{anomk4} are mapped into each other, again in accordance with the duality \eref{zetachimap}.

Similarly to the above discussion, the two-group symmetries of the $\O(2N)^+_{4K}$ gauge theory with $N_f$ flavours given by \eref{twogroupO2N} are mapped to those of the $\O(2N')^{+}_{-4K}$ gauge theory with $N'= N_f+ 2|K| -N+1$ and $N_f$ flavours. Again, in \eref{twogroupO2N}, the Postnikov classes highlighted in blue are interchanged, whereas each of those denoted in black is mapped to itself under the duality \eref{dualityN3}.  We can establish the same statement for the two-groups symmetries of the $\Spin(2N)_{4K}$ gauge theory with $N_f$ flavours and those of the $\O(2N')^{-}_{-4K}$ gauge theory with $N_f$ flavours.


\subsection{$\SO(2N)_{4 K+2}$ gauge theory and open questions} \label{sec:level4Kp2}
Let us now consider the 3d $\CN=3$ $\SO(2N)_{4K+2}$ gauge theory ($K \in \BZ$) with $N_f$ hypermultiplets in the vector representation.  The global form of the flavour symmetry is $\USp(2N_f)/\BZ_2$, since (1) the dressed monopole operators involve an even number of chiral fields, and (2) the operators that do not carry a magnetic flux transform in representations whose highest weights are multiples of that of the adjoint representation. The index of this theory can be computed as explained in Appendix \ref{app:index}.  Here we report the result with all gauge magnetic fluxes and background flavour magnetic fluxes set to $1/2$.
\bes{ \label{SO2N4kp2}
\begin{tabular}{|c|c|c|c|c|c|}
\hline
$N$ & $N_f$ & $k$ & $\chi$& index \\
\hline
$2$ &&  $2$ & $+1$  & $\frac{\zeta}{2} \left[ x^{N_f}  (1+ \sum_{1\leq i<j \leq N_f} f_i f_j)  + \ldots \right] $ \\
       & $\begin{Bmatrix} \text{even} \\ \text{odd} \end{Bmatrix}$ &         &  $-1$  & $\begin{Bmatrix} 1 \\ i \end{Bmatrix} \times \frac{\zeta^{\frac{1}{2}}}{2} \left[ x^{\frac{2N_f-1}{2}}  \left( \sum_{i=1}^{N_f} f_i  \right) + \ldots \right] $ \\
\hline
$3$ &  & $2$ & $+1$  & $\frac{\zeta^{\frac{3}{2}}}{4} \left[ x^{\frac{3N_f-3}{2}}  \left(\sum_{i=1}^{N_f} f_i +\sum_{1\leq i<j<k \leq N_f} f_i f_j f_k \right)  +\ldots \right]$\\
 & $\begin{Bmatrix} \text{even} \\ \text{odd} \end{Bmatrix}$ &  & $-1$   &  $\begin{Bmatrix} 1 \\ i \end{Bmatrix} \times \frac{\zeta}{4} \left[ x^{\frac{3N_f-4}{2}}  \left(\sum_{1\leq i<j \leq N_f} f_i f_j \right)  +\ldots \right]$\\
\hline
\end{tabular}
}
Observe the presence of odd degrees of polynomials in $(f_1, \ldots, f_{N_f})$ for $\chi=-1$ when $N$ is even, and for $\chi=+1$ when $N$ odd.  These features persist for all Chern-Simons levels $k=4K+2$.  We, however, do not have a good understanding of the anomaly theory that is compatible with duality \eref{dualityN3} and duality map \eref{zetachimap} at present,\footnote{For $N$ even and $N_f$ even, we see from the index that if we turn off $B_1^\CC$ (\ie~ $\chi=+1$), there should be no mixed anomaly involving $B_1^\CM$ and $w_2^f$.  Since the duality maps a theory with $N$ even and $N_f$ even to another theory with $N$ even and $N_f$ even, using Table \eref{anomindex}, we see that there are two possibilities for the anomaly theory in this case: either there should be no anomaly or there is an anomaly $\exp\left( i \pi \int B_1^\CC \cup (B_1^\CC +B_1^\CM) \cup w_1^f \right)$, where indeed upon setting $B_1^\CC=0$ there is no mixed anomaly. Even though we do not have a clear understanding of the index results for $\chi=-1$, this seems to suggest that there is still some anomaly and so we conjecture that the anomaly theory should be $\exp\left( i \pi \int B_1^\CC \cup (B_1^\CC +B_1^\CM) \cup w_1^f \right)$. However, for $N$ odd and $N_f$ even we see from the index that when $B_1^\CC=0$ (\ie~ $\chi=+1$), there is an anomaly involving $B_1^\CM$. This theory is also mapped to a theory with $N$ odd and $N_f$ even under the duality, so we see from Table \eref{anomindex} that the anomalies that are compatible with the $\chi=+1$ index and with the duality are either $\exp\left( i \pi \int B_1^\CM\cup B_1^\CM\cup w_2^f\right)$ or $\exp\left( i \pi \int(B_1^\CM\cup B_1^\CM+B_1^\CC\cup B_1^\CC+B_1^\CM\cup B_1^\CC)\cup w_2^f\right)$. Again we do not have a clear understanding of the index result in \eref{SO2N4kp2} for $\chi=-1$ and so we are not able to identify which one is the correct anomaly theory. Indeed, each anomaly theory in \eref{anomindex} that is mapped to itself under the duality is not compatible with the $\chi=-1$ index. In particular, rows 1, 5 and 7 of Table \eref{anomindex} give an imaginary coefficient in the index for $\chi = -1$ and some value of $\zeta$; however, in \eref{SO2N4kp2}, we see that for $\chi =-1$ there is no imaginary number for both $\zeta =\pm1$.  It is possible that the form of the anomaly theory is more complicated than those presented in Table \eref{anomindex}, or that the prescription of computing the indices \eref{SO2N4kp2} needs to be improved.} and leave this open problem for a future investigation.

\subsection{$\U(1)_k$ gauge theory with $N_f$ hypermultiplets} \label{sec:3dN3U1k}
In this section, we consider the 3d $\CN=3$ $\U(1)_k$ gauge theory with $N_f$ hypermultiplets of charge $q$, where we denote this theory by $\CT^k_{N_f, q}$.  The index of this theory is given by
\bes{ \label{indTkNfq}
\CI_{\CT^k_{N_f, q}}(w, n| \vec{f}, \vec{m}; x) &= \sum_{l \in \BZ +\epsilon(m)} \CF_{\CT_q} (w, n| \vec{f}, \vec{m} | l; x)~, \\
\CF_{\CT^k_{N_f, q}} (w, n| \vec{f}, \vec{m} | l; x) &\equiv w^{l} \oint \frac{dz}{2\pi i z} z^{n+k l}~ \prod_{i=1}^{N_f} \CI^{\frac{1}{2}}_\chi \left(z^q f_i^{-1}; q l -f_i; x \right) \times  \\
& \qquad \qquad     \CI^{\frac{1}{2}}_\chi \left(z^{-q} f_i; -q l + f_i; x \right)~, 
}
where $w$ is the fugacity for the $\U(1)_w$ topological symmetry of the theory, and $f_{1,\ldots, N_f}$ are the fugacities for the flavour symmetry algebra $\su(N_f)$.  Using the similar argument as in \eref{qgauging}, we see that if we gauge a $\BZ_q$ subgroup of the $\U(1)_w$ magnetic symmetry of $\CT^k_{N_f, q=1}$ leads to the $\CT^{k q^2}_{N_f, q}$ theory.\footnote{Here $q^2$ comes from the redefinition $z' =z^q$ and $l'=lq$, each of which gives a factor of $q$.}

For $k=0$, the global form of the $\su(N_f)$ flavour symmetry is $\SU(N_f)/\BZ_{N_f}$.  This can be seen from the power series in $x$ of the index $\CI_{\CT^k_{N_f, q}}(w, n =0 | \vec{f}, \vec{m} = \vec 0; x)$ that the dependence of $f_i$ are in terms of the characters of representations of $\SU(N_f)$ of the form $[m,0,\ldots,0,m]$, where $[1,0,\ldots,0,1]$ is the adjoint representation of $\su(N_f)$, for some $m$.  The operators associated with these terms transform trivially under the $\BZ_{N_f}$ centre of $\SU(N_f)$.  Moreover, there is no term involving a product of $w$ and $f_i$ in the index, since the bare monopole operators are gauge neutral and so they are not dressed by the chiral fields.  The global symmetry of the $\CT^{k=0}_{N_f, q}$ theory is therefore $\SU(N_f)/\BZ_{N_f} \times \U(1)_w$, for general $N_f$ and $q$.  There is also a mixed anomaly between $\SU(N_f)/\BZ_{N_f}$ and $\U(1)_w$ which can be detected by considering $\CF_{\CT^{k=0}_{N_f, q}} (w, n| \vec{f}, \vec{m} | l; x)$ with $n=0$, $\vec{m} = \left(\frac{1}{N_f}, \ldots, \frac{1}{N_f}, - \frac{N_f-1}{N_f} \right)$ and $l =\frac{1}{qN_f}$, where the result contains the prefactor $w^{\frac{1}{q N_f}}$.   This mixed anomaly is characterised by the anomaly theory
\bes{
\exp \left(\frac{2\pi i}{N_f} \int w_2^f \cup c_1^w \,(\text{mod }N_f) \right)~,
}
where $w_2^f$ is the generalised 2nd Stiefel-Whitney class that obstructs the $\SU(N_f)/\BZ_{N_f}$ bundles to the $\SU(N_f)$ bundles, and $c_1^w$ is the first Chern-class associated with the $U(1)_w$ topological symmetry.  For the special case of $q=1$, this is in agreement with \cite[(2.6)]{Genolini:2022mpi} and \cite[(7.86)]{Bhardwaj:2022dyt}. For $N_f=2$ we recover \eqref{eq:anomq2}.

Let us now consider the case of $k>0$.   For convenience, let
\bes{
s = \GCD(k, q)~, \qquad \kappa = k/s~, \qquad \mathfrak{q} = q/s~.
}
In addition to the above discussion, there are terms involving the product between $w^{\mathfrak{q}}$ and $w^{-\mathfrak{q}}$ and the characters of the representations $\Sym^\kappa \, \mathbf{N_f} = [\kappa, 0, \ldots, 0]$ and $\Sym^\kappa\, \bar{\mathbf{N_f}}= [0, \ldots, 0, \kappa]$ of $\su(N_f)$. These corresponds to the monopole operators $V_{+\mathfrak{q}}$ and $V_{-\mathfrak{q}}$, carrying the $\U(1)$ gauge charges $-k \mathfrak{q}$ and $k \mathfrak{q}$, dressed by $Q^\kappa$ and $\tQ^\kappa$, where $Q$ and $\tQ$ are the chiral fields carrying gauge charges $+q$ and $-q$.  Since $q \kappa - k \mathfrak{q} = 0$, these dressed monopoles indeed carry zero gauge charge.  The faithful global (non-$R$) symmetry is $[\SU(N_f) \times \U(1)_w]/\BZ_\kappa$.  The $\BZ_\kappa$ action can be seen from the index \eref{indTkNfq} as follows. If we put $\vec f = \left(\frac{1}{N_f}, \ldots, \frac{1}{N_f}, - \frac{N_f-1}{N_f} \right)$ then the argument $q l -f_i \in \BZ$ implies that $l \equiv \frac{1}{qN_f} \, (\mod \, q)$. From the factor $z^{n+k l}$, we must have $n$, which is the background magnetic flux of the $\U(1)_w$ topological symmetry, being $-\frac{k}{qN_f}= -\frac{\kappa}{\mathfrak{q} N_f} \, (\mod \, 1)$.\footnote{If $\frac{\kappa}{\mathfrak{q} N_f} \in \BZ$, then $n$ can be turned off, and the global (non-$R$) symmetry is $\SU(N_f)/\BZ_{\kappa} \times \U(1)_w$.}

Let us assume that $k>0$ and that $q$ divides $k$. Here, $s=\GCD(k,q)=q$, $\mathfrak{q}=1$ and $\kappa =k/q$.  The analysis is very similar to that of \cite[Section 6.2]{Benini:2018reh}, where $\kappa$ in this paper is denoted by $\ell$ in that reference. This theory has a $\BZ_q$ one-form symmetry, since the $\BZ_q$ subgroup of the $\BZ_k$ symmetry acts non-trivially on the Wilson lines that are not screened by the matter fields. Suppose further that $\GCD(\kappa, q) >1$.\footnote{Recall that this is indeed the case for the $\CT^{k q^2}_{N_f, q}$ theory with $q>1$, which arises from gauging the $\BZ_q$ magnetic symmetry of the $\CT^k_{N_f, q=1}$ theory.} It was pointed out in \cite[(6.6)]{Benini:2018reh} that there is a {\it two-group symmetry} between the $\BZ_q$ one-form symmetry and the $[\SU(N_f) \times \U(1)_w]/\BZ_{\kappa = k/q}$ flavour symmetry.  The correponding Postnikov class is $\mathrm{Bock}(w_2^{(\kappa)})$ where $w_2^{(\kappa)}$ is the generalised 2nd Stiefel-Whitney class that obstructs the lift of the $[\SU(N_f) \times \U(1)_w]/\BZ_\kappa$ bundles to the $\SU(N_f) \times \U(1)_w$ bundles, and $\mathrm{Bock}$ is the Bockstein homomorphism associated with the short exact sequence:
\bes{
0 \, \longrightarrow \, \BZ_{q} \, \longrightarrow \, \BZ_{k} \, \longrightarrow \, \BZ_{\kappa = k/q} \, \longrightarrow \, 0~.
}
If $\GCD(q, \kappa)=1$, then $\BZ_{k} = \BZ_q \times \BZ_\kappa$ and the exact sequence splits; in which case, the two-group symmetry is trivial.  

Let us focus on the cases in which $q$ is equal to $1$ or $2$, where the above results can be reconciled with those of the $\SO(2)_k$ and $\Spin(2)_k $ gauge theories.  Indeed, the theory $\CT^k_{2N_f, q=1}$ can also be described as the $\SO(2)_k$ gauge theory with $N_f$ hypermultiplets in the vector representation.  Gauging the $\BZ_2$ subgroup of the $\U(1)_w$ topological symmetry of this theory leads to the $\CT^{4k}_{2N_f, q=2}$ theory, which can also be described by the $\Spin(2)_k$ gauge theory with $N_f$ hypermultiplets in the vector representation.  As we discussed earlier, when $k$ is 0 mod 4, the latter theory has a two-group symmetry between the $\BZ_2$ one-form symmetry and the $\USp(2N_f)/\BZ_2$ symmetry.  This is also the case for the $\CT^{4k}_{2N_f, q=2}$ theory, where it follows from the above discussion if we set the background field for the $\U(1)_w$ topological symmetry to zero and restrict the background field for the flavour symmetry to be in $\USp(2N_f)/\BZ_2$.

%

\section{Non-invertible symmetries in the ABJ-type theories} \label{sec:noninvertibleABJ}
In this section, we investigate mixed anomalies involving a one-form symmetry and two zero-form symmetries, and the non-invertible symmetries arising from gauging appropriate symmetries that participate in such anomalies.

\subsection{$\mathfrak{so}(2N)_{2k}$ gauge algebra with $N_f$ adjoints}
In \cite{Kaidi:2021xfk} it was shown that a 3d theory with a non-invertible symmetry can be constructed starting from a theory with one $\mathbb{Z}_2$ one-form symmetry and two $\mathbb{Z}_2$ zero-form symmetries with the mixed anomaly
\bes{ \label{anomaly_noninv}
\exp \left( i \pi \int B_2\cup B_1^{(1)}\cup B_1^{(2)} \right)~,
}
where $B_2$ is the background field for the one-form symmetry and $B_1^{(i)}$ for $i=1,2$ are those for the two zero-form symmetries. In such a situation gauging two of the symmetries doesn't break the third one as one would naively expect from the anomaly, instead it was explained in \cite{Kaidi:2021xfk} that it makes it non-invertible. 

Suppose first that we gauge the two zero-form symmetries. Then the anomaly \eqref{anomaly_noninv} would make the codimension two topological defect associated to the one-form symmetry non-gauge invariant, but the gauge invariance can be restored by dressing it with a suitable two-dimensional TQFT. The fusion rules for the dressed defect can then be deduced from the tensor product property of the TQFT and typically result in the defect not being invertible. In particular, it was shown in \cite{Kaidi:2021xfk} that for an anomaly of the form \eqref{anomaly_noninv} the dressed defect $\mathcal{N}(M_1)$ for the one-form symmetry obeys the non-group-like fusion rule
\bes{ \label{fusion}
\mathcal{N}(M_1)\times \mathcal{N}(M_1)=\prod_{i=1}^2\left(1+L^{(i)}(M_1)\right)~,
}
which implies that it does not admit an inverse. In \eqref{fusion}, $L^{(i)}(M_1)=\exp\left(i\pi\oint_{M_1} b_1^{(i)}\right)$ and $b_1^{(i)}$ are the dynamical fields for the two gauged zero-form symmetries, so $L^{(i)}(M_1)$ are the codimension two topological defects that are associated to the two one-form symmetries that are dual to the zero-form symmetries that were gauged which obey ordinary group-like fusion fules and are invertible.

Similarly, we can gauge the one-form symmetry and one of the two zero-form symmetries to make the second one non-invertible. Again it was shown in \cite{Kaidi:2021xfk} that dressing the codimension one topological defect associated to the remaining zero-form symmetry with a three-dimensional TQFT makes it gauge invariant, but at the price of having it obey the non-group-like fusion rule
\bes{ \label{fusion2}
\mathcal{N}^{(1)}(M_2)\times \mathcal{N}^{(1)}(M_2)=\frac{1+W(M_2)}{|H^0(M_2,\mathbb{Z}_2)|}\sum_{M_1\in H_1(M_2,\mathbb{Z}_2)}L^{(2)}(M_1)~,
}
which implies that it does not admit an inverse. In \eqref{fusion2}, $W(M_2)=\exp\left(i\pi\oint_{M_2} b_2\right)$, $L^{(2)}(M_1)=\exp\left(i\pi\oint_{M_1} b_1^{(2)}\right)$ and $b_2$, $b_1^{(2)}$ are the dynamical fields for the gauged one-form and zero-form symmetries respectively, so $W(M_2)$ and $L^{(2)}(M_1)$ are the codimension one and two topological defects that are associated to the zero-form and one-form symmetries that are dual to the one-form and zero-form symmetries respectively that were gauged and they obey ordinary group-like fusion fules and are invertible.

One of the examples considered in \cite{Kaidi:2021xfk} (see also \cite[Section 8.4]{Bhardwaj:2022yxj}) is the non-supersymmetric $\SO(2N)_{2k}$ theory with $2N_f$ adjoint real scalars for $N$, $k$ and $N_f$ even. This theory has a $\mathbb{Z}_2$ one-form symmetry coming from the centre of the gauge group which acts trivially on the matter fields and the monopole operators and two $\mathbb{Z}_2$ zero-form symmetries which are the magnetic and the charge conjugation symmetries. Between these symmetries there is an anomaly which is precisely of the form \eqref{anomaly_noninv} \cite{Cordova:2017vab}\footnote{For $N$ and $k$ odd there are additional terms in the anomaly theory that are not linear in $B_1^{\mathcal{M}}$ and $B_1^{\mathcal{C}}$ \cite{Cordova:2017vab} so the procedure of \cite{Kaidi:2021xfk} is not applicable in a straightforward way. It would be interesting to investigate what happens in such cases.}
\bes{ \label{anomaly_noninv_ABJ}
\exp \left( i \pi \int B_2\cup B_1^{\mathcal{M}}\cup B_1^{\mathcal{C}} \right)~.
}
This then authomatically implies that the $\Pin^+(2N)_{2k}$ theory with $2N_f$ adjoint real scalars, which is obtained by gauging both the magnetic and the charge conjugation symmetries, has a non-invertible one-form symmetry for $N$, $k$ and $N_f$ even. Similarly, the $\Spin(2N)_{2k}/\mathbb{Z}^{[1]}_2 = \mathrm{Sc}(2N)_{2k}\,\, \text{or}\,\, \mathrm{Ss}(2N)_{2k}$\footnote{Here we follow the notation of \cite{Aharony:2013hda}. The theories $\mathrm{Sc}(2N)$ and $\mathrm{Ss}(2N)$ for $N$ even are obtained by gauging respectively the $\mathbb{Z}_2^C$ and the $\mathbb{Z}_2^S$ one-form symmetries of the $\Spin(2N)$ theory, where $\mathbb{Z}_2^C$ comes from the centre symmetry of $\Spin(2N)$ acting on one spinor representation and the vector representation while $\mathbb{Z}_2^S$ acts on the other spinor and again the vector. Since $\mathrm{Sc}(N)$ is related to $\mathrm{Ss}(N)$ by the $\BZ_2$ outer-automorphism of $\Spin(N)$, we will not discuss them separately.  In fact, there is another way to obtain this theory.  As pointed out in \cite[Footnote 29]{Cordova:2017vab}, the zero-form magnetic symmetry of the $\SO(2N)/\BZ_2$ gauge theory is extended to $\BZ_4$.  If we gauge a $\BZ_2$ subgroup of the latter, we arrive at  the $\Spin(2N)_{2k}/\mathbb{Z}^{[1]}_2 = \mathrm{Sc}(2N)_{2k}\,\, \text{or}\,\, \mathrm{Ss}(2N)_{2k}$ gauge theory, as required.  Indeed, the commutant of $\BZ_2$ in $\BZ_4$ is identified with the $\BZ_2$ zero-form magnetic symmetry in the resulting theory.  Moreover, in this theory, there is a mixed anomaly between the $\BZ_2$ one-form symmetry, arising from gauging the zero-form symmetry, and the $\BZ_2$ magnetic symmetry.} ({\it resp.} $\O(2N)_{2k}/\mathbb{Z}^{[1]}_2 = \mathrm{PO}(2N)_{2k}$) gauge theory with $2N_f$ adjoint real scalars, which are obtained by gauging the $\mathbb{Z}^{[1]}_2$ one-form symmetry and the magnetic ({\it resp.} charge conjugation) symmetry, has a non-invertible zero-form symmetry for $N$, $k$ and $N_f$ even.

Before discussing how the procedure of \cite{Kaidi:2021xfk} can be applied to the ABJ theory, we will first do the propedeutic exercise of showing how it is possible to use the superconformal index to detect the anomaly \eqref{anomaly_noninv_ABJ} in the supersymmetric version of this theory.\footnote{The definition of the magnetic and the charge conjugation symmetries is slightly different in the supersymmetric and in the non-supersymmetric case, see for example \cite{Aharony:2013kma,Cordova:2017vab}, but also in the supersymmetric case we find the anomaly \eqref{anomaly_noninv_ABJ}.} Namely, we consider the 3d $\mathcal{N}=3$ $\SO(2N)_{2k}$ theory with $N_f$ adjoint hypers for $N$ and $k$ even.\footnote{One could also consider the case with $\mathcal{N}=2$ supersymmetry, which we expect to behave similarly.} In order to detect the anomaly \eqref{anomaly_noninv_ABJ} we compute the index of the theory in which the one-form symmetry has been gauged, which amounts to including monopole sectors with half-integer flux, and refined with fugacities $\chi$ and $\zeta$ for the zero-form symmetries.

Computing the index for low $N$, $k$ and $N_f$ and for $\chi=+1$ we find\footnote{For simplicity we don't turn on any fugacity for the continuous symmetry acting on the matter fields.} \footnote{For $k=0$ we consider $N_f\ge 2$ in order for the theory not to be bad in the sense of \cite{Gaiotto:2008ak}.}
\bes{
N=2,k=0,N_f=2:\quad &1+(21+\zeta)x+\mathcal{O}(x^2)\\
N=2,k=0,N_f=3:\quad &1+42x+40x^{\frac{3}{2}}+(832+\zeta)x^2+\mathcal{O}(x{^\frac{5}{2}})\\
N=2,k=2,N_f=1:\quad &1+6x+13x^2-2(3+\zeta)x^{\frac{5}{2}}+\mathcal{O}(x^3)~.
}
If we instead compute the same indices but for $\chi=-1$ we find
\bes{
N=2,k=0,N_f=2:\quad &1+8x^{\frac{1}{2}}+(32+\zeta^{\frac{1}{2}})x+\mathcal{O}(x^{\frac{3}{2}})\\
N=2,k=0,N_f=3:\quad &1+12x^{\frac{1}{2}}+72x+280x^{\frac{3}{2}}+(789-\zeta^{\frac{1}{2}})x^{2}+\mathcal{O}(x^{\frac{5}{2}})\\
N=2,k=2,N_f=1:\quad &1+4x^{\frac{1}{2}}+(8-\zeta^{\frac{1}{2}})x+\mathcal{O}(x^{\frac{3}{2}})\\
N=4,k=0,N_f=2:\quad &1+8 x^{\frac{1}{2}}+42 x+164 x^{\frac{3}{2}}+539 x^2+1564 x^{\frac{5}{2}}+4296 x^3\\
&+11552 x^{\frac{7}{2}}+31248 x^4+84316 x^{\frac{9}{2}}+225352 x^5\\
&+589792 x^{\frac{11}{2}}+(1511597+\zeta^{\frac{1}{2}})x^6+\mathcal{O}(x^{\frac{13}{2}})~.
}
The half-integer powers of the fugacity $\zeta$ for the magnetic symmetry when $\chi=-1$ signal that the theory has the anomaly \eqref{anomaly_noninv_ABJ}. 

Although computing the index for higher values of $N_f$ is feasible, increasing $N$ and $k$ is computationally more demanding. Nevertheless, we can find a monopole operator that for any even $N$ and $k$ would give a contribution to the index with half-integer power of $\zeta$, which confirms the presence of the anomaly \eqref{anomaly_noninv_ABJ} for any even $N$ and $k$. If we consider the monopole with flux $\left(\frac{1}{2},\cdots,\frac{1}{2},0\right)$, where the last flux is zero since we are considering the case $\chi=-1$ \cite{Aharony:2013kma}, would give the following contribution to the index:\footnote{This can be easily deduced by looking at the overall factor in front of the q-Pochhammers that appear in the integrand of the integral expression of the index.}
\bes{
\zeta^{\frac{N-1}{2}}x^{\frac{N(N-1)}{2}}\prod_{a=1}^{N-1}z_a^{k}~,
}
where $z_a$ are the $\mathfrak{so}(2N)$ gauge fugacities which we parametrize such that the character of the adjoint representation under which the matter fields transform is
\bes{
\chi_{\text{adj.}}^{\mathfrak{so}(2N)}=N+\sum_{a<b}^Nz_az_b+z_az_b^{-1}+z_a^{-1}z_b+z_a^{-1}z_b^{-1}~,
}
with the last fugacity being set to $z_N=1$ and $z_N^{-1}=-1$ for $\chi=-1$ \cite{Aharony:2013kma}. The presence of the gauge fugacities $z_a$ in the monopole contribution signals that this is not a gauge invariant operator. Nevertheless, when $k$ is even we can dress it with the matter fields to make it gauge invariant and the contribution to the index of the resulting operator would have a half-integer power of $\zeta$ only for $N$ even. 
Hence, the anomaly \eqref{anomaly_noninv_ABJ} is indeed present for any $N$ and $k$ even and for arbitrary $N_f$. This in turn implies, following the general analysis of \cite{Kaidi:2021xfk}, that the 3d $\mathcal{N}=3$ $\Pin(2N)_{2k}$ theory with $N_f$ adjoint hypermultiplets, obtained by gauging both the magnetic and the charge conjugation symmetries \cite{Aharony:2013kma}, has a non-invertible one-form symmetry whose topological defect satisfies the fusion rule \eqref{fusion} if $N$ and $k$ are even. Similarly, the $\Spin(2N)_{2k}/\mathbb{Z}^{[1]}_2 = \mathrm{Sc}(2N)_{2k}\,\, \text{or}\,\, \mathrm{Ss}(2N)_{2k}$ ({\it resp.} $\O(2N)_{2k}/\mathbb{Z}^{[1]}_2 = \mathrm{PO}(2N)_{2k}$) gauge theory with $N_f$ hypermultiplets in the adjoint representation, which are obtained by gauging the $\mathbb{Z}^{[1]}_2$ one-form symmetry and the magnetic ({\it resp.} charge conjugation) symmetry, have a non-invertible zero-form symmetry whose topological defect satisfies the fusion rule \eqref{fusion2} for $N$ and $k$ even.

\subsection{ABJ theories of the orthosymplectic type}
Let us now consider the Aharony-Bergman-Jafferis (ABJ) theories \cite{Aharony:2008gk} with gauge algebra $\mathfrak{so}(2N)_{2k}\times\mathfrak{usp}(2M)_{-k}$ and two bifundamental half-hypermultiplets. We will assume throughout this subsection that $k$ is even\footnote{We remark that for $k$ odd, we encounter the same problem as in \eref{sec:level4Kp2}, namely we cannot find the anomaly that is compatible with the generalised level-rank duality \cite{Aharony:2008gk, Honda:2017nku}.}.

First we consider the $\SO(2N)_{2k}\times \USp(2M)_{-k}$ version of the theory.  This has a $\mathbb{Z}^{[1]}_2$ one-form symmetry coming from the diagonal combination of the centres of the two gauge nodes\footnote{Both $\SO(2N)$ and $\USp(2N)$ groups have a $\BZ_2$ centre; see \cite[Table 3]{Cordova:2017vab}.  Since the Wilson line in the bifundamental representation is screened by a matter field, an anti-diagonal combination of $\BZ_2 \times \BZ_2$ is absent. The bare monopole operators, despite not being gauge invariant, are also uncharged under the diagonal combination and so we are left with one $\BZ_2$ one-form symmetry, which is denoted here by $\mathbb{Z}^{[1]}_2$.} (see \cite{Bergman:2020ifi, Beratto:2021xmn}).  There are also the $\mathbb{Z}_2$ zero-form magnetic and charge conjugation symmetries.\footnote{In this section we turn off any background field for the $\SO(3)_f$ flavour symmetry of the ABJ theory. It would be interesting to understand what is its fate after the gaugings that we are going to perform as a consequence of the anomalies \eqref{anomABJw2A}-\eqref{anomABJw2B}, in particular if it also becomes non-invertible.} We will show momentarily using the index that for $N$ even and arbitrary $M$ the theory has exactly the anomaly \eqref{anomaly_noninv_ABJ}. This in turn implies that the $\Pin(2N)_{2k}\times \USp(2M)_{-k}$ variant of the theory has a non-invertible one-form symmetry whose topological defect satisfies the fusion rule \eqref{fusion} for $N$ even. Similarly, the $(\Spin(2N)_{2k}\times \USp(2M)_{-k})/\mathbb{Z}^{[1]}_2$\footnote{Here the gauged $\mathbb{Z}^{[1]}_2$ one-form symmetry is the diagonal combination of either the $\mathbb{Z}_2^S$ or the $\mathbb{Z}_2^C$ centre symmetries of $\Spin(2N)$ and the $\mathbb{Z}_2$ centre symmetry of $\USp(2N)$.} and the $(\O(2N)_{2k}\times \USp(2M)_{-k})/\mathbb{Z}^{[1]}_2$\footnote{Here the gauged $\mathbb{Z}^{[1]}_2$ one-form symmetry is the diagonal combination of the centre symmetries of $\O(2N)$ and $\USp(2N)$.} variants of the theory have a non-invertible zero-form symmetry, which is the charge conjugation symmetry for $\Spin$ and the magnetic symmetry for $\O$, whose topological defects satisfy the fusion rule \eqref{fusion2} for $N$ even.

In order to detect the anomaly, we compute the index of the theory $[\SO(2N)_{2k}\times \USp(2M)_{-k}]/\mathbb{Z}^{[1]}_2$ where the one-form symmetry is gauged and refined with the fugacities $\chi$ and $\zeta$ for the zero-form symmetries. For low $N$, $M$ and $k$ and for $\chi=+1$ we find
\bes{
N=2,M=1,k=2:\quad &1+x+(5+6\zeta)x^2+\mathcal{O}(x^3)\\
N=M=2,k=2:\quad &1+x+(16+11\zeta)x^2+\mathcal{O}(x^3)\\
N=2,M=1,k=4:\quad &1+x+4x^2-4x^3+(4+9\zeta)x^4+\mathcal{O}(x^4)\\
N=M=2,k=4:\quad &1+x+10x^2+(31+24\zeta)x^3+\mathcal{O}(x^4)~,
}
while for $\chi=-1$ we have
\bes{
N=2,M=1,k=2:\quad &1+(1-3\zeta^{\frac{1}{2}})x+\mathcal{O}(x^2)\\
N=M=2,k=2:\quad &1+x+(-2+3\zeta^{\frac{1}{2}}+5\zeta)x^2+\mathcal{O}(x^3)\\
N=2,M=1,k=4:\quad &1+x+(2-5\zeta^{\frac{1}{2}})x^2+\mathcal{O}(x^3)\\
N=M=2,k=4:\quad &1+x-2x^2+4x^3+(4+3\zeta^{\frac{1}{2}}+9\zeta)x^4+\mathcal{O}(x^5)~.
}
Similarly to the example of the previous subsection, the presence of states carrying $\zeta^{\frac{1}{2}}$ indicates that the theory has the anomaly \eqref{anomaly_noninv_ABJ}.

Again computing the index for higher $N$, $M$ and $k$ becomes computationally challenging, but we can construct a gauge invariant dressed monopole operator with half-integer flux that for $\chi=-1$ gives a contribution with a half-integer power of $\zeta$ for $N$ even and if either $M$ is odd and $k$ abritrary or $M$ is even and $k$ is odd. Indeed, the monopole with flux $\left(\frac{1}{2},\cdots,\frac{1}{2},0;\frac{1}{2},\cdots,\frac{1}{2}\right)$, where the first $N$ entries correspond to the $\mathfrak{so}(2N)$ flux and the last $M$ to the $\mathfrak{usp}(M)$ flux, would give the following contribution to the index:
\bes{ \label{monopolecontr}
\zeta^{\frac{N-1}{2}}\prod_{a=1}^{N-1}\left(z^{\mathfrak{so}}_a\right)^{k}\prod_{i=1}^M\left(z^{\mathfrak{usp}}_i\right)^{-k}~,
}
where $z_a^{\mathfrak{so}}$ are the $\mathfrak{so}(2N)$ gauge fugacities and $z_i^{\mathfrak{usp}}$ the $\mathfrak{usp}(2M)$ ones which we parametrize such that the character of the bifundamental representation of $\mathfrak{so}(2N)\times\mathfrak{usp}(2M)$ under which the matter fields transform is
\bes{ \label{charbifund}
\chi_{\text{bifund.}}^{\mathfrak{so}(2N)\times\mathfrak{usp}(2M)}=\left[\sum_{a=1}^Nz_a^{\mathfrak{so}}+\left(z_a^{\mathfrak{so}}\right)^{-1}\right]\left[\sum_{i=1}^Mz_i^{\mathfrak{usp}}+\left(z_i^{\mathfrak{usp}}\right)^{-1}\right]~.
}
and again we set $z_N^{\mathfrak{so}}=1$ and $\left(z_N^{\mathfrak{so}}\right)^{-1}=-1$ for $\chi=-1$. The presence of the gauge fugacities $z_a^{\mathfrak{so}}$, $z_i^{\mathfrak{usp}}$ for the two gauge nodes in the monopole contribution \eref{monopolecontr} signals that this is not a gauge invariant operator. Nevertheless, we can try to dress it with the matter fields to make it gauge invariant. This can be done by choosing terms in \eref{charbifund} to cancel the gauge fugacities in \eref{monopolecontr}.  If $N$ and $M$ have opposite parity, the choice can be made in such a way that they do not involve $z^{\so}_N$ for any $k$.  For example, for $N=2$ and $M=3$, these terms can be chosen as follows: $(z^{\so}_1)^{-1} z^{\usp}_1$, $z^{\so}_1 z^{\usp}_2$, and $(z^{\so}_1)^{-1} z^{\usp}_3$; upon raising to the power $k$, the product of these terms cancel the gauge contribution in \eref{monopolecontr}, as required.  If $N$ and $M$ have the same parity and $k$ is even, such a choice that does not involve $z^{\so}_N$ can be made again.  For example, for $N=M=k=2$, we can choose from the {\it square} of \eref{charbifund} the following terms: $(z^{\so}_1)^{-2} (z^{\usp}_1)^2$ and $ (z^{\usp}_2)^2$, which cancel precisely the gauge fugacities in \eref{monopolecontr}.  However, if $N$ and $M$ have the same parity and $k$ is odd, such a choice must involve $z^{\so}_N$ or $(z^{\so}_N)^{-1}$.  For example, for $N=M=2$ and $k=1$,we can choose from \eref{charbifund} the following terms: $(z^{\so}_1)^{-1} z^{\usp}_1$ and $(z^{\so}_2)^{\pm 1} (z^{\usp}_2)$.  Since for $\chi=-1$ we have to set $z_N^{\mathfrak{so}}=1$ and $\left(z_N^{\mathfrak{so}}\right)^{-1}=-1$, the two contributions cancel against each other. Hence, if $N$ and $M$ have the same parity and $k$ is odd, we cannot obtain the gauge invariant dressed monopole operators.  However, since we focus on $k$ even, we will not consider the latter case; thus, the gauge invariant dressed monopole operators can be formed independent of the parity of $N$ and $M$.

The contribution to the index of the resulting dressed monopole operator has a half-integer power of $\zeta$ only for $N$ even, so the anomaly \eqref{anomaly_noninv_ABJ} is present for any $M$, assuming that $k$ is even.
This in turn implies, following again the analysis of \cite{Kaidi:2021xfk}, that the $\Pin(2N)_{2k}\times \USp(2M)_{-k}$ ABJ theory, with $N$ and $k$ even, has a non-invertible one-form symmetry and similarly the $(\Spin(2N)_{2k}\times \USp(2M)_{-k})/\mathbb{Z}^{[1]}_2$ and the $(\O(2N)_{2k}\times \USp(2M)_{-k})/\mathbb{Z}^{[1]}_2$ theories have a non-invertible zero-form symmetry for $N$ and $k$ even.

\section{Two-groups in the ABJ theories of the orthosymplectic type} \label{sec:twogroupsABJ}
Let us now consider another interesting aspect of the ABJ theories \cite{Aharony:2008gk}, namely the two-group symmetries.  We focus again on those with gauge algebra $\mathfrak{so}(2N)_{2k}\times\mathfrak{usp}(2M)_{-k}$, with $k$ even.  The two bifundamental half-hypermultiplets transform under the flavour symmetry algebra $\su(2)_f$ as a doublet.  In fact, it can be checked using the superconformal index that there is no operator transforming under the $\BZ_2$ centre of $\SU(2)_f$, and so the global form of this flavour symmetry is in fact $\SO(3)_f$. Another way to see this is by observing that a transformation by the $\mathbb{Z}_2$ centre of the flavour symmetry can always be reabsorbed by a gauge transformation for the $\mathbb{Z}_2$ centre of either the gauge groups. It is crucial to remark that, for generic $N$, $M$ and $k$, this theory has $\CN=5$ supersymmetry and the $\SO(3)_f$ symmetry is a subgroup of the $\SO(5)$ R-symmetry.\footnote{Note that the amount of supersymmetry can be larger than $\CN=5$, for example, it can be $\CN=6$ for $k=1$.  However, since we restrict ourselves to $k$ even, we will only consider the theories with $\CN=5$ supersymmetry.}

We examine the mixed anomalies between the magnetic symmetry, the charge conjugation symmetry and the $\SO(3)_f$ flavour symmetries. In the following, we denote the background fields for the former two by $B_1^\CM$ and $B_1^\CC$ respectively, and denote by $w_2^f$ the 2nd Stiefel-Whitney class that obstructs the lifting the $\SO(3)_f$ bundles to the $\SU(2)_f$ bundles.  Let us first summarise the main results and then provide the evidence and reasons later. 
\ben
\item Let us take $N= 2n+1$. The $\SO(4n+2)_{2k} \times \USp(2M)_{-k}$ theory with $k$ even (and $M$ arbitrary) has a mixed anomaly between the $\BZ_2$ zero-form magnetic symmetry, the $\BZ_2$ zero-form charge conjugation symmetry and the $\SO(3)_f$ symmetry given by the anomaly theory
\bes{ \label{anomABJw2A}
\exp \left( i \pi \int B_1^\CM \cup  (B_1^\CM + B_1^\CC ) \cup w_2^f \right)~.
}
We remark that this is consistent with the third line of \eref{anomk4}.
\item Let us take $N= 2n$. The $\SO(4n)_{2k} \times \USp(2M)_{-k}$ theory with $k$ even has a mixed anomaly given by the anomaly theory
\bes{ \label{anomABJw2B}
\exp \left( i \pi \int B_1^\CM \cup   B_1^\CC \cup w_2^f \right) \exp \left( i \pi \int B_2\cup B_1^{\mathcal{M}}\cup B_1^{\mathcal{C}} \right),
}
where the first factor is consistent with the first line of \eref{anomk4} and the second factor comes from \eref{anomaly_noninv_ABJ} and was discussed extensively in the previous section.
\een

These mixed anomalies involving only the zero-form symmetries can be seen from the indices that can be computed in following way.  For the charge conjugation fugacity $\chi=+1$ ({\it resp.} $\chi=-1$), all of the magnetic fluxes of the $\so(2N)$ gauge algebra are set to $1/2$ ({\it resp.} except the last one $m_{N}$ is set to zero and the gauge fugacities $z_N$ and $z^{-1}_N$ are set to $1$ and $-1$ respectively), those for the $\usp(2M)$ gauge algebra are set to $0$, and those for the $\SO(3)_f$ flavour symmetry are set to $1/2$.\footnote{This is because we can choose to reabsorbe a non-trivial transformation for the $\mathbb{Z}_2$ centre of the flavour symmetry by a gauge transformation for the $\mathbb{Z}_2$ centre of the $\SO(2N)$ gauge group only.}
\bes{
\begin{tabular}{|c|c|c|c|c|}
\hline
$N$ & $M$ & $k$ & $\chi$ & index \\
\hline
$2$ & $1$ &  $2$ & $+1$ & $\frac{1}{2} \zeta \left[ x^3 (f^2 + f^4) - x^4  \left( 1+ f^2 \right) +\ldots \right] $ \\
       &        &         & $-1$ & $ -\zeta^{\frac{1}{2}} \left[ x^2 - x^3  f^2 -x^4(1-f^2+f^4) +\ldots \right] $ \\
\hline
$2$ & $2$ &  $2$ & $+1$ & $\frac{1}{2} \zeta \left[ x^5 (f^2 + f^4) - x^6  \left( 1+ f^2 \right) +\ldots \right] $ \\
       &        &         & $-1$ & $ \frac{1}{2}\zeta^{\frac{1}{2}} \left[ x^4 - x^5  f^2 -x^6(1-f^2+f^4) +\ldots \right] $ \\
\hline  \hline     
$3$ & $2$ & $2$ & $+1$ & $\frac{1}{4} \zeta^{\frac{3}{2}} \left[ x^6(1+f^2) + \ldots \right]$ \\
       &        &         & $-1$ & $ \frac{1}{4} \zeta \left[ x^5 f^4 - x^6 f^2 +\ldots \right] $ \\
\hline
$3$ & $3$ & $2$ & $+1$ & $\frac{1}{4} \zeta^{\frac{3}{2}} \left[ x^9(1+f^2) + \ldots \right]$ \\
       &        &         & $-1$ & $ -\frac{1}{4} \zeta \left[ x^8 f^4 - x^9 f^2 +\ldots \right] $ \\
\hline 
\end{tabular}
}
The mixed anomalies between zero-form symmetries can be immediately written down as described in \eref{anomindex}.  This is consistent with the statements in \eref{anomABJw2A} and \eref{anomABJw2B}.


An immediate consequence of the mixed anomalies is the two-group symmetries that are present upon gauging various symmetries.  Below we use the same notation as in \eref{twogroupSpin2N} and \eref{twogroupO2N}.
\bi
\item Gauging the zero-form magnetic symmetry leads to the ABJ theory $\Spin(2N)_{2k} \times \USp(2M)_{-k}$.  For $N$ odd and $k$ even ($M$ arbitrary), the two-group symmetry is characterised by
\bes{
\delta B_2^\CM = \mathrm{Bock}(w_2^f) + B_1^\CC \cup w_2^f~.
}
For $N$ even and $k$ even, the two-group is characterised by
\bes{
\delta B_2^\CM = B_1^\CC \cup (w_2^f + B_2)~.
}
\item Gauging the zero-form charge conjugation symmetry leads to the ABJ theory $\O(2N)_{2k} \times \USp(2M)_{-k}$. For $N$ odd and $k$ even ($M$ arbitrary), the two-group symmetry is characterised by
\bes{
\delta B_2^\CC = B_1^\CM \cup w_2^f~.
}
For $N$ even and $k$ even, the two-group is characterised by
\bes{
\delta B_2^\CC  = B_1^\CM \cup (w_2^f + B_2)~.
}
\ei

To summarise some of the results of this section and the previous one, we saw that starting from the ABJ theory of the $\SO$-type for $N$ and $k$ even which has a non-trivial mixed anomaly, we can go via suitable gaugings either to variants with non-invertible symmetries ($\Pin$, $\Spin/\mathbb{Z}_2^{[1]}$ or $\O/\mathbb{Z}_2^{[1]}$) or to some with two-group symmetries ($\Spin$ or $\O$). This is analogous to similar findings that were pointed out in \cite{Bhardwaj:2022yxj} for the non-supersymmetric $\mathfrak{so}(4N)$ pure gauge theory in arbitrary $d$ dimensions.

Hitherto we have discussed mixed anomalies and two-group symmetries that involve the $\SO(3)_f$ flavour symmetry.  However, since $\SO(3)_f$ is a subgroup of a larger $\SO(5)$ $R$-symmetry, it is natural to ask whether the whole $R$-symmetry participates in such mixed anomalies or two-group symmetries. It is not clear to us how the superconformal index can be used to answer this question.  We hope to investigate and address this issue in future work.

\section{Conclusions}
\label{sec:concl}

In this paper we investigated discrete mixed anomalies in a variety of 3d $\mathcal{N}\geq 3$ theories using the superconformal index as our main tool. These include the $\mathrm{U}(1)_k$ gauge theory with $N_f$ hypermultiplets of charge $q$, the $T(\SU(N))$ theory of Gaiotto-Witten, the theories with $\mathfrak{so}(2N)_{2k}$ gauge algebra with hypermultiplets in the vector representation, and variants of the Aharony-Bergman-Jafferis (ABJ) theory with the orthosymplectic gauge algebra. We then exploited this knowledge to argue, following various constructions available in the literature, that different global variants of these theories obtained by gauging some anomalous symmetries possess non-invertible or two-group symmetries.

There are several open questions and possible directions for future investigations that one might pursue. Among the open questions, it would be interesting to understand the anomalies of the $\SO(2N)_{2k}$ gauge theories with vector matter and of the $\SO$ variant ABJ theories for $k$ odd and how to reconcile them with the known dualities that they enjoy. Another interesting question is whether the two-group involving the $\SO(3)_f$ flavour symmetry of the ABJ theory that we found can be extended to a two-group for the entire $R$-symmetry, of which $\SO(3)_f$ becomes part due to the supersymmetry enhancement at low energies. 

One possible line of future research is to try to apply the same analysis also to other theories for which we can compute the superconformal index, possibly with less supersymmetry. The index indeed becomes particularly useful when studying mixed anomalies of more complicated quiver gauge theories, such as the 3d $\mathcal{N}=2$ theories arising from the compactifications of 5d SCFTs on Riemann surfaces with flux \cite{Sacchi:2021afk,Sacchi:2021wvg,Sacchi:2023rtp,wip}.

Finally, it would be interesting to check our results about mixed anomalies, two-group symmetries and non-invertibles symmetries in the ABJ theories from the holographic perspective, along the lines of for example \cite{Witten:1998wy,Bergman:2020ifi,Bah:2020uev,Apruzzi:2021phx,Bergman:2022otk,vanBeest:2022fss,Antinucci:2022vyk}.

\acknowledgments
We thank Luca Viscardi for the collaboration during the early stages of the work. We would like to express our gratitude to a number of people for several useful discussions: Fabio Apruzzi, Pietro Benetti Genolini, Cyril Closset, Simone Giacomelli, Ho Tat Lam, Shu-Heng Shao, Alberto Zaffaroni, and Gabi Zafrir. N.~M. gratefully acknowledges support from the Simons Center for Geometry and Physics, Stony Brook University, at which part of this research project was conducted. He is also indebted to Raffaele Savelli for the kind hospitality during the completion of this project. MS is partially supported by the ERC Consolidator Grant \#864828 “Algebraic Foundations of Supersymmetric Quantum Field Theory (SCFTAlg)” and by the Simons Collaboration for the Nonperturbative Bootstrap under grant \#494786 from the Simons Foundation.

\appendix

\section{3d supersymmetric index conventions} \label{app:index}

In this appendix we give a brief review of the 3d supersymmetric index \cite{Bhattacharya:2008zy,Bhattacharya:2008bja, Kim:2009wb,Imamura:2011su, Kapustin:2011jm, Dimofte:2011py}. This is also to explain our conventions, which are the same as those used in \cite{Beratto:2021xmn}.

The index can be expressed as a matrix integral of the form
\be\label{indexpartitionfunction}
\mathcal{I}(\{\vec{f},\vec{m}\})\,=\,\sum_{\vec{n}}\frac{1}{|\mathcal{W}_{\vec{n}}|}\oint_{\mathbb{T}^{\text{rk}G}}\prod_{i=1}^{\text{rk}G}\frac{\udl{u_i}˘}{2\pi i u_i} \mathcal{Z}_{\text{cl}}(\{\vec{u},\vec{n}\})\,\mathcal{Z}_{\text{vec}}(\{\vec{u},\vec{n}\})\,\mathcal{Z}_{\text{mat}}(\{\vec{u},\vec{n}\};\{\vec{f},\vec{m}\})\,.
\ee
In this expression, $\vec{u}$ are the gauge fugacities living in the Cartan of the gauge group $G$ and $\vec{n}$ the corresponding magnetic fluxes living in the co-weight lattice of $G$. Hence, the summation over magnetic fluxes is sensible to the global structure of the gauge group. The integration contour is taken to be the unit circle $\mathbb{T}$ for each integration variable and the prefactor $|\mathcal{W}_{\vec{n}}|$ is the dimension of the Weyl group of the residual gauge symmetry in the monopole background with flux $\vec{n}$.  Finally, $\{\vec{f},\vec{m}\}$ denote possible fugacities and fluxes for global symmetries.

The integrand of \eqref{indexpartitionfunction} has three types of contributions. First, we have the classical contribution $\mathcal{Z}_{\text{cl}}$, which can consist of Chern--Simons (CS) interactions and, when the gauge group contains some abelian factor, FI interactions. For example, for a $\U(N)$ gauge group it takes the form
\bea
\mathcal{Z}^{\U(N)}_{\text{cl}}(\{\vec{u},\vec{n}\})=\prod_{i=1}^Nu_i^{kn_i}w^{n_i}\,,
\eea
where $k$ is the CS level and $w$ is the fugacity associated with the $\U(1)_w$ zero-form topological symmetry. In the main text we consider also $\USp(2N)$, $\SO(2N)$ and $\SO(2N+1)$ gauge groups, for which the classical contribution is
\bea
\mathcal{Z}_{\text{cl}}^{\USp(2N)}(\{\vec{u},\vec{n}\})&=\prod_{i=1}^Nu_i^{2kn_i}\nn\\
\mathcal{Z}^{\SO(2N+\epsilon)}_{\text{cl}}(\{\vec{u},\vec{n}\})&=\prod_{i=1}^Nu_i^{2kn_i}\zeta^{n_i}\,,
\eea
where for compactness we denoted $\SO(2N+\epsilon)$ for $\epsilon=0,1$.
Moreover, $\zeta$ is the fugacity for the zero-form topological symmetry, which is $\U(1)_{\zeta}$ for $\SO(2)\cong \U(1)$, while it is a $\mathbb{Z}_2^\mathcal{M}$ for $\SO(2N)$ and $\SO(2N+1)$ with $N>1$ so in these cases we have the condition $\zeta^2=1$.

Then we have the contribution $\mathcal{Z}_{\text{vec}}$ of a 3d $\mathcal{N}=2$ vector multiplet, which takes the following generic form:
\be\label{indvect}
\mathcal{Z}_{\text{vec}}(\{\vec{u},\vec{n}\})\,=\,\prod_{\alpha\in\mathfrak{g}}x^{-\frac{|\alpha(\vec{n})|}{2}}(1-(-1)^{\alpha(\vec{n})}\vec{u}^{\alpha}x^{|\alpha(\vec{n})|})\,,
\ee
where $\alpha$ are the roots of the gauge algebra $\mathfrak{g}$ of the gauge group $G$ and we are using the short-hand notations
\be
\vec{u}^\ga=\prod_{i=1}^{\text{rk}G}u_i^{\ga_i},\quad \ga(\vec{n})=\sum_{i=1}^{\text{rk}G}\ga_in_i,\quad|\alpha(\vec{n})|=\sum_{i=1}^{\text{rk}G}\ga_in_i\,.
\ee
Explicitly for the groups of main interest in this paper we have
\bea\label{induspvect}
\mathcal{Z}_{\text{vec}}^{\USp(2N)}(\{\vec{u},\vec{n}\})&=\prod_{i=1}^Nx^{-2|n_i|}\prod_{s=\pm1}(1-(-1)^{2s\,n_i}u_i^{2s}x^{2|s\,n_i|})\nn\\
&\times\prod_{i<j}^Nx^{-|n_i+n_j|-|n_i-n_j|}\prod_{s_1,s_2=\pm1}(1-(-1)^{s_1n_i+s_2n_j}u_i^{s_1}u_j^{s_2}x^{|s_1n_i+s_2n_j|})\nn\\
\mathcal{Z}_{\text{vec}}^{\SO(2N+\epsilon)}(\{\vec{u},\vec{n}\},\chi=+1)&=\left(\prod_{i=1}^Nx^{-|n_i|}\prod_{s=\pm1}(1-(-1)^{s\,n_i}u_i^sx^{|s\,n_i|})\right)^\epsilon\nn\\
&\times\prod_{i<j}^Nx^{-|n_i+n_j|-|n_i-n_j|}\prod_{s_1,s_2=\pm1}(1-(-1)^{s_1n_i+s_2n_j}u_i^{s_1}u_j^{s_2}x^{|s_1n_i+s_2n_j|})\,.
\eea
For $\SO(2N+\epsilon)$ we also have a discrete zero-form charge conjugation symmetry $\mathbb{Z}_2^\mathcal{C}$ whose corresponding fugacity in the index we denote by $\chi$ and the above expressions hold for $\chi=+1$. For $\chi=-1$ we have to set $u_N=+1$, $u_N^{-1}=-1$ and $n_N=0$ when $\epsilon=0$, while when $\epsilon=-1$ we have \cite{Hwang:2011qt,Hwang:2011ht,Aharony:2013kma}
\bea
\mathcal{Z}_{\text{vec}}^{\SO(2N+1)}(\{\vec{u},\vec{n}\},\chi=-1)&=\prod_{i=1}^Nx^{-|n_i|}\prod_{s=\pm1}(1+(-1)^{s\,n_i}u_i^sx^{|s\,n_i|})\nn\\
&\times\prod_{i<j}^Nx^{-|n_i+n_j|-|n_i-n_j|}\prod_{s_1,s_2=\pm1}(1-(-1)^{s_1n_i+s_2n_j}u_i^{s_1}u_j^{s_2}x^{|s_1n_i+s_2n_j|})\,.
\eea

Finally, we have the contribution $\mathcal{Z}_{\text{mat}}$ of matter fields which come into 3d $\mathcal{N}=2$ chiral multiplets. The contribution of a chiral with $R$-charge $r$ and transforming under a $\U(1)$ symmetry with fugacity and flux $u$ and $n$ respectively is
\bes{\label{indchir}
\CZ_{\text{chir}} (u; n ; r) = (x^{1-r} u^{-1})^{|n|/2} \prod_{p=0}^\infty \frac{1-(-1)^n u^{-1} x^{|n|+2-r +2p}}{1-(-1)^n u x^{|n|-r +2p}}~.
}
When $r$ is taken to be the superconformal $R$-charge, then the supersymmetric index coincides with the superconformal index of the SCFT to which the gauge theory flows in the IR. The full contribution to the index of a set of chirals transforming in representations $\mathcal{R}_G$ and $\mathcal{R}_F$ of the gauge and the flavour symmetry respectively and with $R$-charge $r$ is
\bea
\mathcal{Z}_{\text{mat}}(\{\vec{u},\vec{n}\};\{\vec{f},\vec{m}\},r)&=\prod_{\rho_G\in\mathcal{R}_G}\prod_{\rho_F\in\mathcal{R}_F}\CZ_{\text{chir}} (\vec{u}^{\rho_G}\vec{f}^{\rho_F}; \rho_G(\vec{n})+\rho_F(\vec{m}) ; r)\,,
\eea
where $\rho_G$ and $\rho_F$ are the weights of $\mathbf{R}_G$ and $\mathbf{R}_F$ respectively. Notice that a chiral in the adjoint representation of the gauge group and with $R$-charge 1 gives a trivial contribution to the index. Thus, the index factor for a 3d $\mathcal{N}=4$ vector multiplet, which decomposes into an $\mathcal{N}=2$ vector and an $\mathcal{N}=2$ chiral, actually coincides with the one of an $\mathcal{N}=2$ vector multiplet \eqref{indvect}. An example of a matter field that we encountered in the main text is an $\mathcal{N}=4$ hypermultiplet in the bifundamental of $\SO(2N+\epsilon)\times \USp(2M)$, where in the $\SO(2N+\epsilon)$ gauge theory $\USp(2M)$ is a flavour symmetry while in the ABJ theory it is a gauge symmetry. Its contribution to the index is (we take the superconformal $R$-charge $r=1/2$)
\bea
&\mathcal{Z}^{\SO(2N+\epsilon)\times \USp(2M)}_{\text{mat}}(\{\vec{u},\vec{n}\};\{\vec{z},\vec{m}\};\chi=+1)=\left(\prod_{j=1}^M\prod_{s=\pm1}\CZ_{\text{chir}} (z_j^s; s\,m_j; 1/2)\right)^\epsilon \nn\\
&\qquad\qquad\qquad\qquad\qquad\qquad\qquad\times\prod_{i=1}^N\prod_{j=1}^M\prod_{s_1,s_2=\pm1}\CZ_{\text{chir}} (u_i^{s_1}z_j^{s_2}; s_1n_i+s_2m_j ; 1/2)\,,
\eea
where $\{\vec{u},\vec{n}\}$ are the $\SO(2N+\epsilon)$ fugacities and fluxes, while $\{\vec{z},\vec{m}\}$ are the $\USp(2M)$ fugacities and fluxes.
The last expression holds again only for $\chi=+1$. The correct contribution for $\chi=-1$ in the case $\epsilon=0$ is obtained by setting $z_N=1$, $z_N^{-1}=-1$ and $m_N=0$, while when $\epsilon=1$ we have the compact expression for generic $\chi$ \cite{Hwang:2011qt,Hwang:2011ht,Aharony:2013kma}
\bea
&\mathcal{Z}^{\SO(2N+1)\times \USp(2M)}_{\text{mat}}(\{\vec{u},\vec{n}\};\{\vec{z},\vec{m}\};\chi)=\prod_{j=1}^M\prod_{s=\pm1}\CZ_{\text{chir}} (\chi\,z_j^s; s\,m_j; 1/2) \nn\\
&\qquad\qquad\qquad\qquad\qquad\qquad\qquad\times\prod_{i=1}^N\prod_{j=1}^M\prod_{s_1,s_2=\pm1}\CZ_{\text{chir}} (u_i^{s_1}z_j^{s_2};s_1n_i+s_2m_j ; 1/2)\,.
\eea

To conclude, let us explain how to obtain the indices of the theories with gauge algebras $\mathfrak{so}(N)$ and $\mathfrak{usp}(2M)$ for different choices of the global structure of the gauge group. Let us start from the simplest case of $\mathfrak{usp}(2M)$. If we take the gauge group to be $\USp(2M)$ then we have to sum over integer magnetic fluxes $\vec{n}\in\mathbb{Z}^M$, while if we take it to be $\USp(2M)/\mathbb{Z}_2$ then we have to sum over half-integer magnetic fluxes $\vec{n}\in(\mathbb{Z}/2)^M$. In the latter case one can also introduce a fugacity $g$ obeying $g^2=1$ for the $\mathbb{Z}_2$ magnetic symmetry so that the monopoles with integer magnetic flux carry $g^0=1$ while those with half-integer magnetic flux carry $g^1$.

In the case of $\mathfrak{so}(N)$ there several possible global variants, which depend on the value of $N$. In the main text we only consider the case of $\mathfrak{so}(2N)$ and in particular we compute the index for the groups $\SO(2N)$ and $\SO(2N)/\mathbb{Z}_2$. These work similarly to the $\mathfrak{usp}(2M)$ case, that is for $\SO(2N)$ we sum over integer fluxes while for $\SO(2N)/\mathbb{Z}_2$ we sum over half-integer fluxes. There are also other groups, some of which we also encountered in the main text but we didn't need to compute their index explicitly. For example, $\mathfrak{so}(N)$ for any $N$ admits the variants $\Spin(N)$, $\O(N)$ and $\Pin(N)$. These are obtained from the $\SO(N)$ theory by gauging the magnetic and the charge conjugation symmetries. At the level of the index, the gauging is implemented by summing over all the possible values of their fugacities $\zeta=\pm1$ and $\chi=\pm1$ and dividing by the dimension of the corresponding symmetry group \cite{Aharony:2013kma}
\bea
\mathcal{I}_{\Spin(N)}(\chi)&=\frac{1}{2}\left(\mathcal{I}_{\SO(N)}(\zeta=+1;\chi)+\mathcal{I}_{\SO(N)}(\zeta=-1;\chi)\right)\nn\\
\mathcal{I}_{\O(N)^+}(\zeta)&=\frac{1}{2}\left(\mathcal{I}_{\SO(N)}(\zeta;\chi=+1)+\mathcal{I}_{\SO(N)}(\zeta;\chi=-1)\right)\nn\\
\mathcal{I}_{\O(N)^-}(\zeta)&=\frac{1}{2}\left(\mathcal{I}_{\SO(N)}(\zeta;\chi=+1)+\mathcal{I}_{\SO(N)}(-\zeta;\chi=-1)\right)\nn\\
\mathcal{I}_{\Pin(N)}&=\frac{1}{2}\left(\mathcal{I}_{\Spin(N)}(\chi=+1)+\mathcal{I}_{\Spin(N)}(\chi=-1)\right)\nn\\
&=\frac{1}{4}\left(\mathcal{I}_{\SO(N)}(\zeta=+1;\chi=+1)+\mathcal{I}_{\SO(N)}(\zeta=-1;\chi=+1)\right.\nn\\
&\left.\qquad+\mathcal{I}_{\SO(N)}(\zeta=+1;\chi=-1)+\mathcal{I}_{\SO(N)}(\zeta=-1;\chi=-1)\right)\,,
\eea
where for simplicity we only specify the dependence on $\zeta$ and $\chi$ which are the only relevant ones for the gauging.

\bibliographystyle{ytphys}
\bibliography{ref}

\end{document}